\documentclass[%
 aip,
 jap,
 amsmath,amssymb, 
 preprint,
]{revtex4-2}

\usepackage[T1]{fontenc}
\usepackage[utf8]{inputenc}
\usepackage{mathptmx}
\usepackage{amsmath}
\usepackage[english]{babel}
\usepackage{graphicx}
\usepackage{booktabs}
\usepackage{dcolumn}
\usepackage{etoolbox}
\usepackage[colorlinks=true, allcolors=blue]{hyperref}

\begin{document}

\title{Thermal Transport in Defective Uranium Nitride: Effects of Point Defects, Anharmonicity, and Electronic Contributions}

\author{Beihan Chen}
 \affiliation{Department of Nuclear Engineering, The Pennsylvania State University, University Park, PA 16802, USA}
 
\author{Marat Khafizov}
\affiliation{Department of Mechanical and Aerospace Engineering, The Ohio State University, Columbus, OH 43210, USA}

\author{Zilong Hua}
\affiliation{Idaho National Laboratory, Idaho Falls, ID 83415, USA}

\author{David H. Hurley}
\affiliation{Idaho National Laboratory, Idaho Falls, ID 83415, USA}

\author{Miaomiao Jin}%
 \email{mmjin@psu.edu}
\affiliation{Department of Nuclear Engineering, The Pennsylvania State University, University Park, PA 16802, USA}%

\begin{abstract}

The impact of point defects on thermal transport in uranium nitride (UN) is investigated using a machine learning interatomic potential combined with Green--Kubo (GK) and normal mode analysis (NMA) methods over 300--1500 K. In pristine UN, temperature-dependent calculations of lattice thermal conductivity reveal that four-phonon scattering is essential yet sufficient to accurately capture high-temperature anharmonic phonon transport, as evidenced by close agreement between GK and ShengBTE calculations including three- and four-phonon processes. In defective systems, all types of point defects significantly reduce thermal conductivity at low temperature. Mode-resolved analysis further shows that interstitial defects introduce new phonon states due to a stronger local strain effect. Notably, the uranium interstitial leads to strong defect-phonon scattering over broad phonon spectrum, while the other point defects produce more selective scattering, with even reduced phonon scattering for some acoustic modes. The optical contribution to thermal conductivity remains nearly constant in the presence of $\mathrm{I}_\mathrm{U}$, but decreases with increasing temperature for pristine and the other defect types. The total thermal conductivity, incorporating electron--phonon coupling and an estimated electronic contribution from the Wiedemann–Franz law, yields excellent agreement with experiment in the pristine system, with electronic contributions dominating thermal transport above $\sim$600 K. Moreover, with defect-electron contribution introduced through a semiclassical electron--defect scattering model, it is found that i) the total conductivity degradation  follows the order of uranium interstitials > uranium vacancies > nitrogen interstitials > nitrogen vacancies, and ii) electron--phonon coupling becomes negligible in defective systems. These results provide a unified understanding of defect-dependent thermal transport in UN.


\end{abstract}

\maketitle

\section{Introduction}

Uranium nitride (UN) has attracted significant attention as an accident-tolerant nuclear fuel and a promising candidate for next-generation reactor systems due to its superior thermophysical properties compared to conventional UO$_2$ fuel \cite{Hayes1990tc,Jones2023,Youinou2014}. In particular, UN exhibits a metallic character and relatively high thermal conductivity, which increases with temperature, enabling more efficient heat removal under reactor operating conditions. 

Extensive experimental efforts have been devoted to measuring the thermal conductivity of UN; however, reported values show noticeable variation due to differences in sample density, fabrication methods, and measurement conditions. Early studies by Speidel and Keller \cite{Speidel1963} and Kikuchi \textit{et al.} \cite{Kikuchi1973} highlighted the strong influence of porosity on thermal conductivity of sintered UN over a broad temperature range. Endebrock \textit{et al.} \cite{Endebrock1964} reported values for fully dense arc-melted UN reaching 16.7 W/mK at 323 K. Subsequent measurements by Nasu and Kikuchi \cite{NASU1968} and Moore \textit{et al.} \cite{MOORE1970} established room-temperature values in the range of 11--14 W/mK for high-density (95\% theoretical density sintered) samples. Comprehensive assessments by Ross \textit{et al.} \cite{Ross1988} and Hayes \textit{et al.} \cite{Hayes1990tc} further consolidated these results into empirical correlations. Most recently, Charatsidou \textit{et al.} have experimentally investigated the thermophysical properties of zirconium-incorporated and ion-irradiated uranium nitride fuel \cite{Charatsidou2026Zr,Charatsidou2026}. Among these datasets, the measurements of Takahashi \textit{et al.} \cite{Takahashi1971} for fully dense UN are widely used for comparison \cite{Miller2024,Miller2025,Valter2015}, and agree with the most recent measurement on a single-crystal UN sample at room temperature \cite{Chen2025}; hence, they were adopted in this work for validation.

From a theoretical perspective, thermal transport in UN arises from both lattice ($\kappa_\mathrm{L}$) and electronic ($\kappa_\mathrm{e}$) contributions. Recent advances in computational methods have enabled detailed investigation of these components. Molecular dynamics (MD) simulations using the Green--Kubo formalism with empirical potentials \cite{Green1954,Kubo1957} and solution to Boltzmann transport calculations based on first-principles-derived force constants have been applied to evaluate $\kappa_\mathrm{L}$ in UN \cite{Kurosaki2000,Szpunar2020}. In parallel, density functional theory (DFT) based approaches have been used to quantify $\kappa_\mathrm{e}$ to determine the total thermal conductivity \cite{Kocevski2023}. More recently, a machine learning interatomic potential (MLIP) was developed for UN and used in conjunction with lattice dynamics and Boltzmann transport calculations to predict $\kappa_\mathrm{L}$, explicitly including four-phonon scattering \cite{Chen2025}. The results demonstrated that higher-order anharmonicity plays a critical role in determining thermal transport at elevated temperatures, and that combining $\kappa_\mathrm{L}$ with $\kappa_\mathrm{e}$ yields excellent agreement with experimental data.

Under irradiation, the accumulation of point defects can significantly degrade thermal conductivity \cite{Hurley2022}. While defect-induced thermal transport degradation has been extensively studied in oxide nuclear fuels \cite{Watanabe2009,Liu2016,Resnick2019,Mitchell2020,Hurley2022} and in other nitride systems such as AlN and GaN \cite{Rounds2018,Li2025,Yang2025}, comparatively limited attention has been devoted to UN. Recent studies have begun to address this gap. Lan \textit{et al.} \cite{Lan2025} investigated vacancy-induced degradation of thermal conductivity in UN using DFT-based Boltzmann transport calculations and demonstrated that uranium vacancies cause a more pronounced reduction in thermal conductivity than nitrogen vacancies. Galvin \textit{et al.} \cite{Galvin2025} further evaluated the electronic thermal conductivity ($\kappa_\mathrm{e}$) of defective UN using DFT with Boltzmann transport equation (DFT-BTE) method \cite{Madsen2018}, considering individual point defects (uranium interstitials $\mathrm{I}_\mathrm{U}$, nitrogen interstitials $\mathrm{I}_\mathrm{N}$, uranium vacancies $\mathrm{V}_\mathrm{U}$, and nitrogen vacancies $\mathrm{V}_\mathrm{N}$) at a concentration of 0.46\%. However, their non-equilibrium MD (NEMD) investigation of lattice thermal conductivity was limited to $\mathrm{I}_\mathrm{N}$ and $\mathrm{V}_\mathrm{U}$, and did not account for electron--phonon coupling effects. As a result, a systematic understanding of defect-dependent lattice and total thermal transport in UN remains to be further improved.

In this work, we present a systematic investigation of defect- and temperature-dependent thermal transport in UN using the previously developed MLIP \cite{Chen2025}, extending prior studies on defect-affected UN thermal conductivity \cite{Galvin2025,Lan2025}. Four types of point defects, i.e. $\mathrm{I}_\mathrm{N}$, $\mathrm{I}_\mathrm{U}$, $\mathrm{V}_\mathrm{N}$, and $\mathrm{V}_\mathrm{U}$, are considered at a concentration of 0.46\%. The lattice thermal conductivity is evaluated using both the Green--Kubo (hereafter abbreviated as GK) and normal mode analysis methods (hereafter abbreviated as NMA), enabling a direct comparison between fluctuation-based and phonon-based descriptions of heat transport. In addition, electron--phonon coupling and electronic thermal conductivity are incorporated to determine the total thermal conductivity and to quantify the relative contributions of different heat transport mechanisms. The lattice thermal conductivity considering phonon scattering only is denoted as $\kappa_{\mathrm{L-ph}}$, while considering both phonon scattering and electron--phonon coupling effect is denoted as $\kappa_{\mathrm{L}}$. Within this framework, $\kappa_\mathrm{L}$ is the quantity evaluated rigorously, while the electronic contribution $\kappa_\mathrm{e}$ is estimated using a semi-classical electrical resistivity model that explicitly includes electron–defect scattering \cite{Zhou2020,Zhou2021}. This work provides a unified view to reveal the roles of higher-order anharmonicity, defect-induced scattering, and electronic contributions in governing thermal transport in UN.

\section{Methods}

\subsection{Atomistic modeling and defect construction}

All MD simulations were performed using a previously developed MLIP for UN, which has demonstrated reliability in evaluating defect energetics and thermal properties \cite{Chen2025}. Equilibrium molecular dynamics (EMD) simulations were carried out using LAMMPS \cite{LAMMPS} with periodic boundary conditions applied in all three directions. $6\times6\times6$ supercells of the rock-salt UN structure (1728 atoms) were used unless otherwise specified.

Point defects, including $\mathrm{I}_\mathrm{U}$, $\mathrm{I}_\mathrm{N}$, $\mathrm{V}_\mathrm{U}$, or $\mathrm{V}_\mathrm{N}$, were introduced into the supercell. Interstitial defects were introduced by placing atoms at tetrahedral interstitial sites, while vacancies were created by removing atoms from their lattice positions. Defect concentration at 0.46\%, defined as the ratio of defect atoms to the total number of atoms ($N_{\mathrm{defect}}/N_{\mathrm{total}}$), was considered. Details on the construction of defective supercells are provided in the Supplementary Materials (SM) Sec. 1.

\subsection{Green--Kubo method}

The $\kappa_\mathrm{L-ph}$ was first evaluated using the GK formalism based on linear response theory \cite{Green1954,Kubo1957}. In this approach, $\kappa_\mathrm{L-ph}$ is obtained from the time integral of the heat current autocorrelation function (HCACF),

\begin{equation}\label{eq_gk}
    \kappa_\mathrm{L-ph} = \frac{1}{k_{B}T^{2}V}  \int_0^\infty  \langle  \mathbf{Q}(t) \mathbf{Q}(0) \rangle dt
\end{equation}

where $k_B$ is the Boltzmann constant, $T$ is the temperature, $V$ is the system volume, and $\mathbf{Q}(t)$ is the microscopic heat current vector, defined as

\begin{equation}
    \mathbf{Q}(t)=\sum_i e_i \mathbf{v}_i + \sum_{i<j} (\mathbf{F}_{ij}\cdot \mathbf{v}_i)\mathbf{r}_{ij},
\end{equation}

where $e_i$ and $\mathbf{v}_i$ are the energy and velocity of atom $i$, respectively, $\mathbf{F}_{ij}$ is the force acting on atom $i$ due to atom $j$, and $\mathbf{r}_{ij}$ is the position vector from atom $j$ to atom $i$.

In the EMD simulations, the system was first equilibrated for 10 ps in the isothermal--isobaric (NPT) ensemble, followed by 10 ps in the canonical (NVT) ensemble. The NVT stage was included as a practical equilibration step to allow the system to re-equilibrate at the fixed equilibrium volume while maintaining the target temperature, thereby reducing transient fluctuations before the subsequent microcanonical (NVE) production run. Production runs were then performed for 400 ps in the NVE ensemble. The heat current was sampled every 0.01 ps, and the HCACF was evaluated with a correlation time of 50 ps. The GK integral was computed numerically over this finite correlation-time window, and the final thermal conductivity was obtained from the converged plateau of the running integral within the 50 ps integration window, as shown in SM Fig. S1. Such 50 ps window was confirmed long enough to converge the conductivity. Simulations were performed over a temperature range of 300--1500 K for both pristine and defective systems, in increments of 200 K. To improve statistical convergence, ten independent simulations were performed for each case using different initial velocity seeds. The final $\kappa_\mathrm{L-ph}$ values were obtained by averaging the HCACF over independent runs and integrating the resulting correlation function. The integration window was selected based on the convergence behavior of the HCACF. 

\subsection{Normal mode analysis}

Alternatively, the $\kappa_\mathrm{L-ph}$ is evaluated within the relaxation time approximation (RTA) \cite{Ladd1986,srivastava2022physics} as
\begin{equation}
    \kappa_\mathrm{L-ph}  = \sum_{n} c_n v_{g,n}^2 \tau_{n-\mathrm{ph}},
\end{equation}
where $c_n$, $v_{g,n}$, and $\tau_n$ are the $n$-th mode-specific heat capacity, group velocity, and relaxation time, respectively. The volumetric heat capacity for each mode is given by \cite{PORTER199753}
\begin{equation}
    c_n = \frac{k_B}{V} \frac{x^2 e^x}{(e^x - 1)^2}, \quad x = \frac{\hbar \omega_n}{k_B T},
\end{equation}

where $V$ is the supercell volume, $\omega_n$ is the phonon frequency.

The phonon relaxation times are extracted by NMA on the EMD simulation trajectory. This method has been demonstrated in previous studies, e.g., by McGaughey \textit{et al.} \cite{McGaughey2004} and Turney \textit{et al}. \cite{Turney2009}. The core idea of this method is to project the atomic trajectories obtained from EMD onto individual phonon modes obtained from lattice dynamics (LD) calculations. The LD analysis is conducted using Phonopy \cite{phonopy,phonopy-3py}, interfaced with LAMMPS and based on the developed MLIP, using ground-state 6$\times$6$\times$6 UN supercells for each system. 
The relaxation time of each phonon mode, $\tau_n$, is determined from the temporal decay of the autocorrelation function of the kinetic energy associated with phonon mode $n$ ($E_n$). The kinetic energy $E_n$ is obtained from the time derivative of the normal-mode coordinate of phonon mode $n$, $\dot{A}_n(t)$, i.e.,

\begin{equation}\label{eq_nma_Em}
E_n(t) = \frac{\dot{A}_n(t)^*\dot{A}_n(t)}{2}  ,
\end{equation}
where $\dot{A}_n(t)$ can be expressed as \cite{Henry2008},

\begin{equation}\label{eq_nma_Am} 
\dot{A}_{n}(t) = \sum_{j}\mathrm{exp}(i \cdot \mathbf{k}\cdot\mathbf{r_j(0)}) \mathbf{e}_{j,n}^* \cdot \mathbf{v}_{j}(t) ,
\end{equation}

where $j$ is the index of atom number, $\mathbf{e}_{j,n}$ is the phonon eigenvector (obtained from LD), and $\mathbf{v}_{j}(t)$ is the atomic velocity vector. The phonon kinetic energy $E_n$ and phonon relaxation time $\tau_n$ have this relation \cite{McGaughey2013}:

\begin{equation}\label{eq_nma_ac}
\mathrm{exp}(-\frac{t}{\tau_{n-\mathrm{ph}}})\mathrm{cos}^{2}(\omega_{n} t) = \frac{ \langle E_{n}(t)E_{n}(0) \rangle } { \langle E_{n}(0)E_{n}(0) \rangle} .
\end{equation}

Atomic trajectories were recorded every 0.01 ps and divided into 50 ps segments to improve the statistical robustness. The modal energy autocorrelation function was calculated separately for each segment, and the resulting autocorrelation functions were averaged over all segments and independent MD runs prior to fitting the phonon relaxation times. The 50 ps segment length was chosen to ensure complete decay of the modal autocorrelation functions while providing sufficient independent samples for statistical averaging. For each system and temperature, ten independent simulations were performed, and the resulting autocorrelation functions were averaged prior to fitting based on Eq.~\eqref{eq_nma_ac}. A representative example of the fitting procedure is shown in SM Fig. S2.

\subsection{Electron--phonon coupling and electronic thermal conductivity}
The electronic contribution to thermal transport in UN is significant at 300--600 K, as discussed in our previous work \cite{Chen2025}. To account for electron--phonon interactions, the phonon relaxation time for mode $n$ obtained from NMA was modified using Matthiessen's rule \cite{Tong2019,Ziman2001},
\begin{equation}
    \tau^{-1}_{n} = \tau_{n-\mathrm{ph}}^{-1} + \tau_{n-\mathrm{ep}}^{-1},
    \label{eq8}
\end{equation}
where $\tau_{n}$ is the effective phonon relaxation time, $\tau_{n-\mathrm{ph}}$ is the intrinsic phonon--phonon relaxation time, and $\tau_{n-\mathrm{ep}}$ is the electron--phonon relaxation time.
The electron--phonon scattering rates were obtained from DFT calculations using the EPW package, consistent with our previous work \cite{Chen2025}. The EPW-derived scattering rates were used only in the phonon-side correction of Eq.~\eqref{eq8} and were not employed to evaluate $\kappa_\mathrm{e}$, because the pristine-cell EPW calculations captured only electron–phonon coupling. To properly evaluate $\kappa_\mathrm{e}$, the semiclassical resistivity model (Eqs.~\eqref{rho_general}, \eqref{eq:rho_eph}, \eqref{eq:rho_ee}, and \eqref{eq:rho_epd}) was used because it folds electron-phonon, electron-electron, and electron-defect scattering into a single effective framework.

The total thermal conductivity is then given by
\begin{align}
    \kappa_\mathrm{L} &= \sum_{n} c_n v_{g,n}^2 \tau_{n}, \\
    \kappa_\mathrm{total} &= \kappa_{\mathrm{L}} + \kappa_\mathrm{e},
\end{align}

where the electronic thermal conductivity, $\kappa_\mathrm{e}$ , is estimated based on Wiedemann-Franz Law:

\begin{equation}
    \kappa_\mathrm{e} = \frac{LT}{\rho} ,
\end{equation}

where $T$ is temperature, $L=2.44\times10^{-8} \mathrm{W\cdot\Omega\cdot K^{-2}}$ is the Lorentz number, and $\rho$ is the total electrical  resistivity. To better estimate the impact of point defects on $\rho$, a semi-classical model used by Zhou \textit{et al.} \cite{Zhou2020,Zhou2021} is adopted, as expressed in Eq.~\eqref{rho_general}. This formulation provides an alternative description of the concave temperature dependence of resistivity relative to the crystal-field model used previously \cite{Samsel,Chen2025}, while explicitly enabling the evaluation of point-defect contributions to $\rho$. We adopt this semi-classical method rather than direct DFT-BTE  \cite{Madsen2018,Galvin2025} estimates of $\kappa_\mathrm{e}$ from defective UN because the latter do not explicitly include electron–defect scattering, which is the dominant defect contribution to $\rho$ \cite{Zhou2020}. The current treatment restores this missing channel through the $\rho_\mathrm{e\text{-}pd}$ term, which derives from an electronic scattering cross section $A_\mathrm{cs}$, the Fermi energy $E_\mathrm{F}$, and the defect volume $V_\mathrm{pd}$.

\begin{equation}
\label{rho_general}
    \rho=\frac{\rho_{\mathrm{e-ph}}+\rho_{\mathrm{e-e}}+\rho_{\mathrm{e-pd}}+\rho_\mathrm{res}}{1+(\rho_{\mathrm{e-ph}}+\rho_{\mathrm{e-e}})/\rho_\mathrm{sat}},
\end{equation}
where
\begin{equation}
    \rho_{\mathrm{e-ph}} = \beta T, 
    \label{eq:rho_eph}
\end{equation}

\begin{equation}
    \rho_{\mathrm{e-e}} = \gamma T^2, 
    \label{eq:rho_ee}
\end{equation}

\begin{equation}
    \rho_{\mathrm{e-pd}} = \frac{x A_\mathrm{cs}}{s V_\mathrm{pd}} \sqrt{\frac{2 E_\mathrm{F}}{m_\mathrm{e}}}.
    \label{eq:rho_epd}
\end{equation}

and $\rho_{\mathrm{sat}}$ and $\rho_{\mathrm{res}}$ denote the saturation resistivity and the residual resistivity, corresponding to the high-temperature and low-temperature limits of the resistivity, respectively. Based on Eq.~\eqref{rho_general}, the resulting $\rho$ is constrained by the unsaturated resistivity term in the numerator and the saturation resistivity $\rho_{\mathrm{sat}}$, i.e., Eq.~\eqref{rho_general} enforces a resistivity that does not exceed the saturation limit at high temperature.

For $\rho_{\mathrm{e-ph}}$ and $\rho_{\mathrm{e-e}}$, the parameters $\beta$ for electron–phonon scattering, $\gamma$ for electron–electron scattering, $\rho_{\mathrm{res}}$ for the residual resistivity, and $\rho_{\mathrm{sat}}$ for the saturation resistivity are fitted using experimental electrical resistivity data from Moore \textit{et al.} \cite{MOORE1970}. The fitting results are $\beta$=1.3$\times$10$^{-1} \ \mu\Omega\cdot{cm} \cdot{K^{-1}}$, $\gamma$=2.1$\times$10$^{-11} \ \mu\Omega\cdot{cm}\cdot{K^{-2}}$, $\rho_{\mathrm{res}}$ = 134 $ \mu\Omega$ and $\rho_{\mathrm{sat}}$ = 254 $ \mu\Omega$. These parameters are fitted once to the pristine experimental resistivity data and then held fixed across all defective systems, so that the defect dependence enters only through the defect-specific term $\rho_{\mathrm{e\text{-}pd}}$. 

It should be noted that the experimental samples of Moore \textit{et al.} \cite{MOORE1970} contain a small fraction (0.45–1.6 vol. \%) of secondary phases, which are not explicitly considered in the present transport model. These microstructural features may introduce additional phonon and electron scattering, contributing to some uncertainty in the fitted transport parameters, but are not expected to alter the overall temperature dependence or the conclusion regarding the dominant role of point-defect scattering.

For $\rho_{\mathrm{e\text{-}pd}}$, the point defect concentration $x$ is 0.46\% in this work. The electronic scattering cross section $A_\mathrm{cs}$ is estimated from the  Slater empirical atomic radius \cite{Zhou2020,Slater1964} as $\pi r^{2}$, where is defined such that the sum of the radii of two atoms forming a bond gives an approximate value of the internuclear distance, and the electron mass is taken as $m_\mathrm{e} = 0.511$ $\mathrm{MeV}/c^{2}$. The quantity $s = 1/(\rho \tau_\mathrm{e})$ ($\tau_\mathrm{e}$ is the electron relaxation time) and the Fermi energy $E_\mathrm{F}$ are estimated from DFT–BTE calculations using VASP \cite{Kresse1996,Kresse1996i} combined with BoltzTraP2  \cite{Madsen2018} on a 3$\times$3$\times$3 pristine UN supercell. The point defect volume $V_\mathrm{pd}$ is estimated from the volume change of a $3 \times 3 \times 3$ supercell upon insertion of a point defect, corresponding to the 0.46\% defect concentration, as evaluated using the developed MLIP \cite{Chen2025}, the results are 10.0 $\text{\AA}^2$ for $\mathrm{V}_\mathrm{N}$, 11.2 $\text{\AA}^2$ for $\mathrm{V}_\mathrm{U}$, 13.1 $\text{\AA}^2$ for $\mathrm{I}_\mathrm{N}$ and 22.5 $\text{\AA}^2$ for $\mathrm{I}_\mathrm{U}$. These defect volumes were obtained from structural relaxations of the defective 3×3×3 supercells using the MLIP, in which both the cell parameters and the internal atomic coordinates were fully optimized, and the defect volume ($V_\mathrm{pd}$) is defined as the change in supercell volume per inserted or removed defect relative to the pristine supercell. The same parameters in Eqs.~\eqref{rho_general}, \eqref{eq:rho_eph}, \eqref{eq:rho_ee}, and \eqref{eq:rho_epd} are used for all defective systems, with the defect dependence entering only through $A_\mathrm{cs}$ and $V_\mathrm{pd}$.

\section{ Results and Discussion }

\subsection{$\kappa_\mathrm{L-ph}$ of pristine UN}

\begin{figure}
    \centering
    \includegraphics[width=0.5\linewidth]{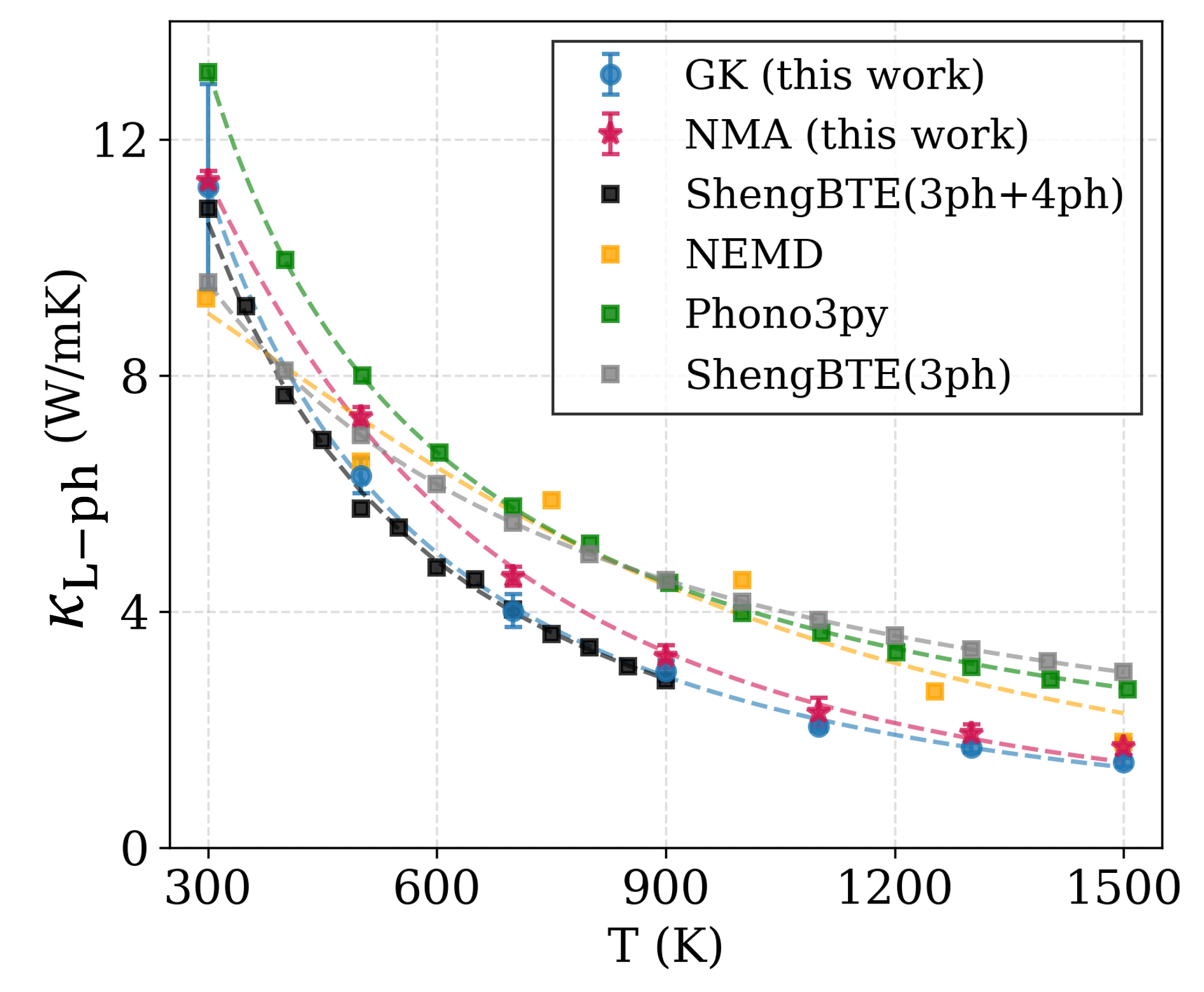}
    \caption{
    $\kappa_\mathrm{L-ph}$ of pristine UN calculated using the GK and NMA methods, compared with ShengBTE results considering four-phonon scattering from previous work \cite{Chen2025} using the same MLIP, NEMD results using ADP \cite{Galvin2025}, Phono3py results based on DFT calculations \cite{Galvin2025}, and ShengBTE results based on DFT calculations considering only three-phonon scattering \cite{Szpunar2020}.  
    }
    \label{fig:1}
\end{figure}

We first analyze the temperature dependence of $\kappa_\mathrm{L-ph}$ of pristine UN to establish a reference for interpreting the defective systems. This comparison provides a direct basis for assessing the consistency of the GK and NMA methods against previous calculations. To quantify the temperature dependence of $\kappa_\mathrm{L-ph}$, we used a modified semi-empirical Klemens--Callaway form \cite{Klemens1955,Klemens1958thermal,Callaway1959}. We retain the Klemens–-Callaway form here because it captures the essential temperature dependence in a compact expression whose parameters can be directly compared across the different computational approaches examined in this study. For pristine UN, the thermal conductivity was fitted using
\begin{equation}
    \kappa_\mathrm{L-ph} = \frac{1}{A+BT+CT^2},
    \label{eq_abc}
\end{equation}

where $A$ is a temperature-independent residual term, while the $BT$ and $CT^2$ terms represent the effective contributions from three-phonon and four-phonon scattering, respectively. The inclusion of the quadratic term is to account for temperature dependence of four-phonon scattering, with the scattering rate $\tau^{-1}_{\mathrm{4ph}}\propto T^2$ \cite{Feng2017,Han2022}. For methods that include only three-phonon scattering, the fitting was performed without the $CT^2$ term. Table~\ref{tab:ABC} summarizes the fitted parameters obtained in this work from the GK and NMA methods, along with parameters extracted by fitting literature data for pristine UN obtained from different computational approaches. These include BTE calculations using ShengBTE packages \cite{Chen2025,Feng2017,Han2022} based on the same MLIP and incorporating both three- and four-phonon scattering; NEMD simulations by Galvin \textit{et al.} \cite{Galvin2025}, who employed an angular-dependent potential (ADP) \cite{Kuksin2016} to calculate the lattice thermal conductivity and combined it with DFT-based Boltzmann transport calculations to estimate the electronic contribution; BTE results using Phono3py \cite{phono3py} combined with DFT calculations; and ShengBTE calculations considering only three-phonon scattering \cite{Szpunar2014}. The corresponding temperature-dependent $\kappa_\mathrm{L-ph}$ data and fitted curves are shown in Fig.~\ref{fig:1}.

\begin{table}[htbp]
    \centering
    \caption{Fitted parameters in Eq.~\eqref{eq_abc} for pristine UN. The parameters $A$, $B$, and $C$ represent the temperature-independent residual term and the effective strengths of three-phonon and four-phonon scattering contributions, respectively. } 
    \label{tab:ABC}
    \begin{tabular*}{\textwidth}{@{\extracolsep{\fill}}llll}
    \hline
    \textbf{Method} & \textbf{$A$} (W$^{-1}$$\cdot$m$\cdot$K)  & \textbf{$B$} (W$^{-1}$$\cdot$m) & \textbf{$C$} (W$^{-1}$$\cdot$m$\cdot$K$^{-1}$) \\ \hline
    GK (this work) & $(1.20\pm0.71)\times10^{-2}$ & $(2.01\pm0.31)\times10^{-4}$ & $(1.87\pm0.28)\times10^{-7}$ \\
    NMA (this work) & $(4.74\pm1.14)\times10^{-2}$ & $(6.31\pm4.99)\times10^{-5}$ & $(2.44\pm0.46)\times10^{-7}$ \\
    ShengBTE(3ph+4ph) \cite{Chen2025} & $(1.44\pm1.41)\times10^{-2}$ & $(2.13\pm0.50)\times 10^{-4}$ & $(1.78\pm0.42)\times10^{-7}$ \\
    NEMD \cite{Galvin2025} & $(9.07\pm3.82)\times10^{-2} $ & $(0.24\pm1.44)\times10^{-4}$ & $(1.39\pm1.13)\times 10^{-7}$ \\
    Phono3py \cite{Galvin2025}& $(2.24\pm0.63)\times 10^{-3}$ & $(2.45\pm0.02)\times 10^{-4}$ & ---\\
    ShengBTE(3ph) \cite{Szpunar2020} & $(4.64\pm0.04)\times10^{-2}$ & $(1.94\pm0.01)\times10^{-4}$ & --- \\ 
    \hline
    \end{tabular*}
\end{table}

The GK results obtained in this work are in close agreement with the ShengBTE calculations that include both three- and four-phonon scattering \cite{Chen2025}, with close values of fitted $A$, $B$, and $C$ parameters. The validity of predicting thermal conductivity via classical MD hinges on the system temperature relative to the Debye temperature ($\Theta_D$), as it intrinsically excites all phonon modes equally according to the classical equipartition theorem. For UN, reported values for $\Theta_D$ range from 181 to 364~K \cite{AbdulHameed2024}. Given that our lowest investigated temperature ($T = 300$~K) lies close to the reported $\Theta_D$ upper limit, the vast majority of phonon modes are sufficiently populated. Consequently, the deviation from quantum Bose-Einstein statistics is expected to be negligible \cite{McGaughey2004}, which is also reflected via the agreement between GK and ShengBTE results.  Given that GK approach implicitly captures all anharmonic scattering processes present in the interatomic potential, whereas ShengBTE explicitly treats phonon scattering within a lattice dynamics framework, such consistency in $B$ and $C$ suggests that four-phonon scattering yields an effective description of anharmonic phonon scattering in pristine UN.  In addition, while the $B$ parameters from GK and ShengBTE(3ph+4ph) are consistent with those obtained from BTE calculations considering only three-phonon scattering \cite{Chen2025,Szpunar2014}, the latter methods systematically overestimate $\kappa_\mathrm{L-ph}$ at elevated temperatures, as shown in Fig.~\ref{fig:1}. Such observations suggest that four-phonon scattering is of significance and can sufficiently describe the high-order phonon transport at elevated temperatures. 

It should be noted that the slight crossing between the ShengBTE calculations including only three-phonon scattering and those including both three- and four-phonon scattering near 300 K is due to the differences in the underlying force constants and computational settings \cite{Szpunar2020}, rather than a physical enhancement of thermal conductivity by four-phonon scattering. Since four-phonon interactions introduce additional phonon scattering channels, their contribution is expected to reduce the lattice thermal conductivity, which is consistently observed at elevated temperatures.

Regarding the NMA method, thermal transport is described within the relaxation time approximation (RTA), where phonon lifetimes are obtained from mode-resolved energy autocorrelation functions extracted from EMD simulations. The lifetimes extracted from MD trajectories inherently include all anharmonic interactions present in the system. Therefore, although the NMA approach here does not distinguish between different scattering orders, it provides an effective description of phonon relaxation. By comparison, the NMA method leads to thermal conductivity values close to GK method, especially at high temperatures, while the fitting yields a smaller $B$ but a larger $C$ compared to the GK results. This indicates that the temperature dependence of $\kappa_\mathrm{L-ph}$ in NMA is effectively shifted toward a stronger higher-order scattering contribution. 

Finally, we note that the NEMD results based on ADP exhibit much higher $\kappa_\mathrm{L-ph}$ in the temperature range of 700--1000 K than values obtained from other methods, leading to a deviation from the polynomial temperature dependence of $1/\kappa_\mathrm{L-ph}$. This behavior is also observed in defective UN systems as reported by Galvin \textit{et al.} \cite{Galvin2025}. It may be attributed to the ADP formalism, which introduces dipole and quadrupole invariants while neglecting higher-order terms \cite{Mishin2005}. 

\subsection{$\kappa_\mathrm{L-ph}$ of defective UN}

\begin{figure}
    \centering
    \includegraphics[width=0.8\linewidth]{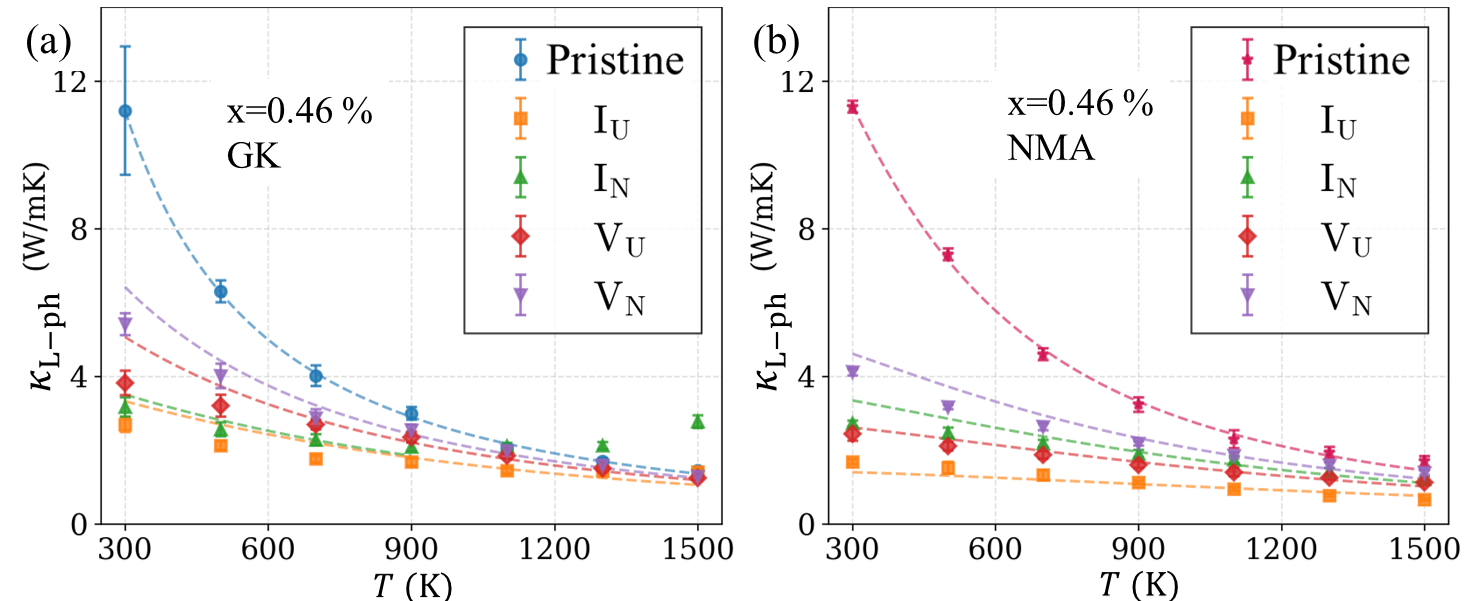}
    \caption{$\kappa_\mathrm{L-ph}$ for pristine and defective UN (0.46\%) obtained from the (a) GK and (b) NMA method from 300 K to 1500 K in increments of 200 K.
    }
    \label{fig:kL_defect}
\end{figure}

The $\kappa_\mathrm{L-ph}$ values of UN containing point defects ($\mathrm{I}_\mathrm{U}$, $\mathrm{I}_\mathrm{N}$, $\mathrm{V}_\mathrm{U}$ or $\mathrm{V}_\mathrm{N}$) at the concentration 0.46\%, calculated using the GK method over the temperature range of 300-1500 K, are shown in Fig.~\ref{fig:kL_defect}(a). At temperatures below $\sim$900 K, all defect types noticeably reduce $\kappa_\mathrm{L-ph}$ relative to pristine UN. The magnitude of degradation follows the order
 $\mathrm{I}_\mathrm{U}$ > $\mathrm{I}_\mathrm{N}$ > $\mathrm{V}_\mathrm{U}$ > $\mathrm{V}_\mathrm{N}$. The stronger effect from interstitials, especially uranium interstitials, is consistent with conventional phonon-defect scattering theory, where larger mass and local structural perturbations introduce stronger phonon scattering centers \cite{Klemens1958thermal,Callaway1959,Walker1963}.

It is interesting to note that for $T > 900$ K, the $\kappa_\mathrm{L-ph}$ of pristine UN and UN containing $\mathrm{I}_\mathrm{U}$, $\mathrm{V}_\mathrm{U}$ and $\mathrm{V}_\mathrm{N}$ converges toward similar values, indicating that phonon--phonon scattering dominates thermal transport in this regime. In contrast, UN containing $\mathrm{I}_\mathrm{N}$ exhibits an anomalous increase in $\kappa_\mathrm{L-ph}$ with increasing temperature. Moreover, this enhancement becomes more pronounced with increasing $\mathrm{I}_\mathrm{N}$ concentration. Additional simulations performed at $T = 1500$ K for multiple $\mathrm{I}_\mathrm{N}$ concentrations confirm that the increase in $\kappa_\mathrm{L-ph}$ scales approximately with defect concentration (see SM Fig. S4). This behavior cannot be explained within a purely phonon-based transport picture. Although nitrogen interstitials exhibit high mobility at elevated temperatures (see mean squared displacement (MSD) analysis presented in the SM Fig. S5), their diffusivity alone is unlikely to contribute to thermal transport at a magnitude comparable to phonon-mediated conduction. Instead, the observed enhancement likely originates from the definition of heat flux within the GK formalism. The microscopic heat current contains a convective term associated with atomic motion. At high temperatures, the increased mobility of $\mathrm{I}_\mathrm{N}$ leads to increasing contributions from this term, and the enhancement positively correlates with $\mathrm{I}_\mathrm{N}$ concentration. As a result, the GK method captures not only phonon-mediated energy transport but also additional contributions arising from defect dynamics, leading to an apparent increase in $\kappa_\mathrm{L-ph}$. In contrast, the NMA framework describes thermal transport in terms of phonon quasiparticles within the relaxation time approximation and therefore does not capture this contribution. Hence, this distinction between GK and NMA becomes critical in systems with mobile defects. We note that this defect-motion contribution is distinct from the coherent, off-diagonal phonon transport neglected by the single-mode RTA discussed below: the former arises from the convective term in the heat current due to atomic migration, whereas the latter is a vibrational channel present in the MD dynamics but discarded in the diagonal NMA-RTA reconstruction.

Comparing $\kappa_\mathrm{L-ph}$ at an equivalent defect concentration (0.46\%), the GK approach yields higher  $\kappa_\mathrm{L-ph}$ values across the entire temperature range for all defect types compared to the NMA-RTA method. This behavior contrasts with the pristine lattice, where NMA predicts slightly higher $\kappa_\mathrm{L-ph}$ values. The underestimation of $\kappa_\mathrm{L-ph}$ by the NMA method in the defective systems can be attributed to two potential factors. The first factor is the limitations of the single-mode RTA \cite{McGaughey2004}. In the presence of spatial disorder and broken translational symmetry, defect scattering introduces significant mode-mode coupling between closely spaced vibrational states, phenomena historically described by Allen and Feldman \cite{allen1989thermal} and more recently unified into a more general theory by Simoncelli \textit{et al.} \cite{simoncelli2019unified}. The NMA-RTA framework neglects these coherent, off-diagonal contributions. In comparison, the GK method evaluates thermal transport directly from full heat current autocorrelations in the time domain. As demonstrated by Lv and Henry \cite{lv2016direct}, this allows GK to inherently capture all degrees of anharmonicity and multi-phonon scattering pathways, circumventing the assumptions built into RTA. A secondary factor contributing to the discrepancy is the numerical uncertainty of extracting precise phonon relaxation times from MD trajectories in defective systems \cite{thomas2010predicting}. In the NMA framework, relaxation times are derived by fitting the temporal decay of modal energy autocorrelation functions (or the frequency-domain linewidth of the spectral energy density). In the presence of point defects, the modal energy autocorrelation functions become noisier and can deviate from the damped oscillatory model of Eq. \eqref{eq_nma_ac} over an extended time range (see SM Fig. S3). Consequently, fitting Eq. \eqref{eq_nma_ac} yields effective phonon lifetimes that are more sensitive to statistical fluctuations. This represents a limitation of the NMA-RTA analysis for defective systems and may contribute to an underestimation, or at least a larger uncertainty, in the NMA-derived $\kappa_\mathrm{L-ph}$.

A further consideration arises in defect configurations where point defects become thermally mobile (in this study, only N interstitial above 900 K). Although the GK formalism remains rigorously valid for evaluating the equilibrium heat-current response, the microscopic heat current then includes a convective component associated with atomic motion. As a result, defect migration can generate additional correlations in the HCACF beyond those due to phonon-mediated heat transport, potentially leading to further overestimation of $\kappa_\mathrm{L-ph}$ \cite{RYU2020109615,Manjunatha2021}. However, it should be noted that this issue is not entirely straightforward: the presence of defect migration may introduce non-negligible complexities, and the sign of its contribution to the HCACF is not trivial to determine a priori, as the migration process likely involves strongly correlated atomic rearrangements. In practice, this means that care is required when interpreting the calculated $\kappa_\mathrm{L-ph}$ as a purely phonon lattice thermal conductivity, particularly in systems where defect mobility is non-negligible.

Consistently, both methods indicate that the degradation in $\kappa_\mathrm{L-ph}$ is  strongest for $\mathrm{I}_\mathrm{U}$ and weakest for $\mathrm{V}_\mathrm{N}$. Within the NMA method, this reflects the varying strengths of phonon scattering cross-sections induced by each defect type (to be discussed next). Following conventional phonon scattering models \cite{Klemens1955, tamura1983isotope}, the major degradation caused by $\mathrm{I}_\mathrm{U}$ is driven by the synergistic effect of a massive local strain field (force-constant variance) and a large mass variance, which strongly scatter a broad spectrum of phonon frequencies. Conversely, $\mathrm{V}_\mathrm{N}$ represents a relatively minor mass and volumetric perturbation. Regarding $\mathrm{I}_\mathrm{N}$ and $\mathrm{V}_\mathrm{U}$, while NMA predicts highly comparable $\kappa_\mathrm{L-ph}$ values for these two defects, GK suggests a stronger degradation induced by $\mathrm{I}_\mathrm{N}$. This discrepancy likely arises because, while $\mathrm{V}_\mathrm{U}$ primarily acts as a mass-difference scatterer \cite{tamura1983isotope}, $\mathrm{I}_\mathrm{N}$ heavily alters local force constants and introduces highly localized, high-frequency vibrational modes \cite{Klemens1955}. These localized modes may enhance the coupling between different vibrational states and give rise to additional transport effects. Although such anharmonic interactions are already embedded in the MD trajectories used in the relaxation time evaluation for NMA, the subsequent single-mode RTA reconstruction neglects off-diagonal modal correlations, whereas these contributions are naturally retained in the GK formalism. However, NMA remains a valuable diagnostic tool for extracting defect-affected phonon relaxation times, particularly their impact on high-frequency modes and low-frequency acoustic windows for each individual defect, as discussed next.
 
\subsection{ Phonon relaxation times}

\begin{figure}
    \centering
    \includegraphics[width=\linewidth]{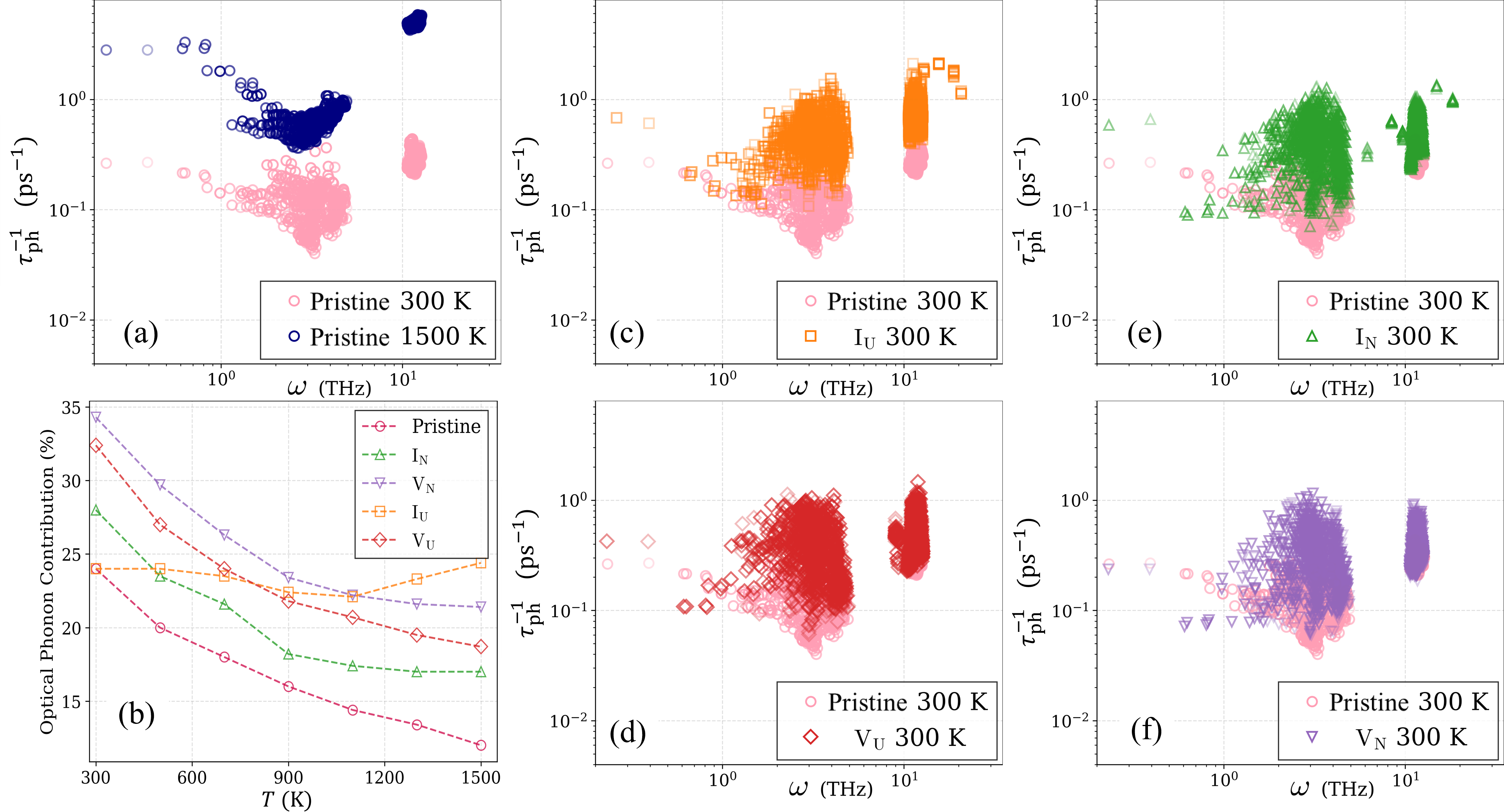}
    \caption{(a) Phonon scattering rates ($\tau_\mathrm{ph}^{-1}$) of pristine UN at 300 K and 1500 K. (b) Optical phonon contributions to $\kappa_\mathrm{L-ph}$ in pristine UN and defect-containing UN systems. (c--f) Comparison of $\tau_\mathrm{ph}^{-1}$ at 300 K between pristine UN and UN containing a 0.46\% concentration of specific point defects:  (c) $\mathrm{I}_\mathrm{U}$,  (d) $\mathrm{V}_\mathrm{U}$,  (e) $\mathrm{I}_\mathrm{N}$, and (f) $\mathrm{V}_\mathrm{N}$, respectively. All $\tau_\mathrm{ph}^{-1}$ were calculated from the NMA method with relaxation time fitting. }
    \label{fig:4}
\end{figure}

The phonon relaxation times for pristine UN and defective UN containing $\mathrm{I}_\mathrm{U}$, $\mathrm{I}_\mathrm{N}$, $\mathrm{V}_\mathrm{U}$ and $\mathrm{V}_\mathrm{N}$ at the concentration of 0.46\% were analyzed by the NMA method at each temperature. The results are presented as phonon scattering rate, $\tau^{-1}_\mathrm{ph}$, in Fig.~\ref{fig:4}(a,c-f). Fig.~\ref{fig:4}(a) compares the $\tau^{-1}_\mathrm{ph}$ of pristine UN at 300~K and 1500~K. As expected, enhanced anharmonicity at elevated temperatures drives an overall increase in $\tau^{-1}_\mathrm{ph}$ across the entire dispersion. The modes exhibiting the longest relaxation times (acting as the major heat carriers) are localized within a mid-frequency acoustic window spanning 3 to 4~THz.  

For defective systems, the scattering rates at 300 K are shown in Fig.~\ref{fig:4}(c--f), together with the pristine reference.  The introduction of interstitial defects ($\mathrm{I}_\mathrm{U}$ and $\mathrm{I}_\mathrm{N}$) leads to new, high-frequency vibrational modes. Notably, the lighter $\mathrm{I}_\mathrm{N}$ defect introduces localized gap states within the phonon bandgap. This is attributed to the stronger strain field associated with interstitial defects \cite{Klemens1955} than that of vacancies. Among all defects, the heavy $\mathrm{I}_\mathrm{U}$ defect induces the most significant increase in $\tau^{-1}_\mathrm{ph}$ across the entire frequency spectrum, stemming from both strong local strain and mass variance. In contrast, for the other defect types ($\mathrm{V}_\mathrm{N}$, $\mathrm{V}_\mathrm{U}$, and $\mathrm{I}_\mathrm{N}$), while an overall increase in $\tau^{-1}_\mathrm{ph}$ is observed, certain low-frequency acoustic modes exhibit reduced scattering rates compared to the pristine system. Such behavior suggests that the presence of defects alters the availability of allowed scattering processes for different phonon modes. This selective phonon survival implies that defect-induced phase-space restriction could lead to certain acoustic modes propagating with reduced resistance.
We further note that the extracted scattering rates do not vanish at low frequency even in pristine UN (Fig. \ref{fig:4}(a)), because they include the intrinsic anharmonic background; this low-frequency plateau is additionally affected by the finite q-point sampling of the 6×6×6 supercell, so the $\omega$ $\rightarrow$ 0 limit should not be interpreted quantitatively. Meanwhile, because the dominant heat-carrying modes in UN lie in the 3–4 THz acoustic window, this low-frequency limitation has only a minor effect on the computed $\kappa_\mathrm{L-ph}$.

To further quantify the role of different phonons, we evaluated the contribution of optical phonons ($\omega > 9$~THz) to the total lattice thermal conductivity $\kappa_\mathrm{L-ph}$ as a function of temperature, as shown in Fig.~\ref{fig:4}(b). In pristine UN, the optical contribution decreases monotonically with increasing temperature and makes a non-negligible contribution to thermal transport in UN, as also reported in UO$_2$ \cite{Pang2013}. This trend is consistent with the increasing importance of anharmonic scattering at elevated temperature, including higher-order phonon processes that are known to strongly affect high-frequency modes \cite{Feng2017}. For most defective systems, a similar decrease is observed; however, the relative optical contribution follows the ordering: $\mathrm{V}_\mathrm{N} > \mathrm{V}_\mathrm{U} > \mathrm{I}_\mathrm{N} > \mathrm{pristine}$ throughout the temperature range. This behavior indicates that defect-induced scattering suppresses acoustic phonons more effectively than optical phonons. Since acoustic modes are the primary heat carriers in pristine UN, their stronger degradation shifts the relative contribution toward higher-frequency optical modes, even though the absolute optical contribution still decreases with temperature. We note that this comparison refers to the relative optical contribution to $\kappa_\mathrm{L-ph}$, not to the absolute defect-induced scattering rates: in absolute terms the high-frequency optical modes still exhibit the largest increases in scattering rate (strongest for $\mathrm{I}_\mathrm{U}$), consistent with the Rayleigh-type enhancement of point-defect scattering at intermediate and high frequencies \cite{Gurunathan2020}.

A qualitatively different behavior is found for the $\mathrm{I}_\mathrm{U}$ system, in which the optical contribution remains nearly constant at approximately 24\% over the entire temperature range. This weak temperature dependence suggests that transport in the presence of $\mathrm{I}_\mathrm{U}$ is dominated by defect-induced structural scattering, such that the additional temperature-dependent anharmonic scattering becomes comparatively less important. In this sense, the $\mathrm{I}_\mathrm{U}$ case appears to move toward a more disorder-dominated transport regime, where the relative branch contributions are constrained more by defect-induced scattering than by intrinsic phonon--phonon interactions \cite{hanus2021thermal}.

\subsection{Electron--phonon coupling and electronic contributions}

\begin{figure}
    \centering
    \includegraphics[width=0.9\linewidth]{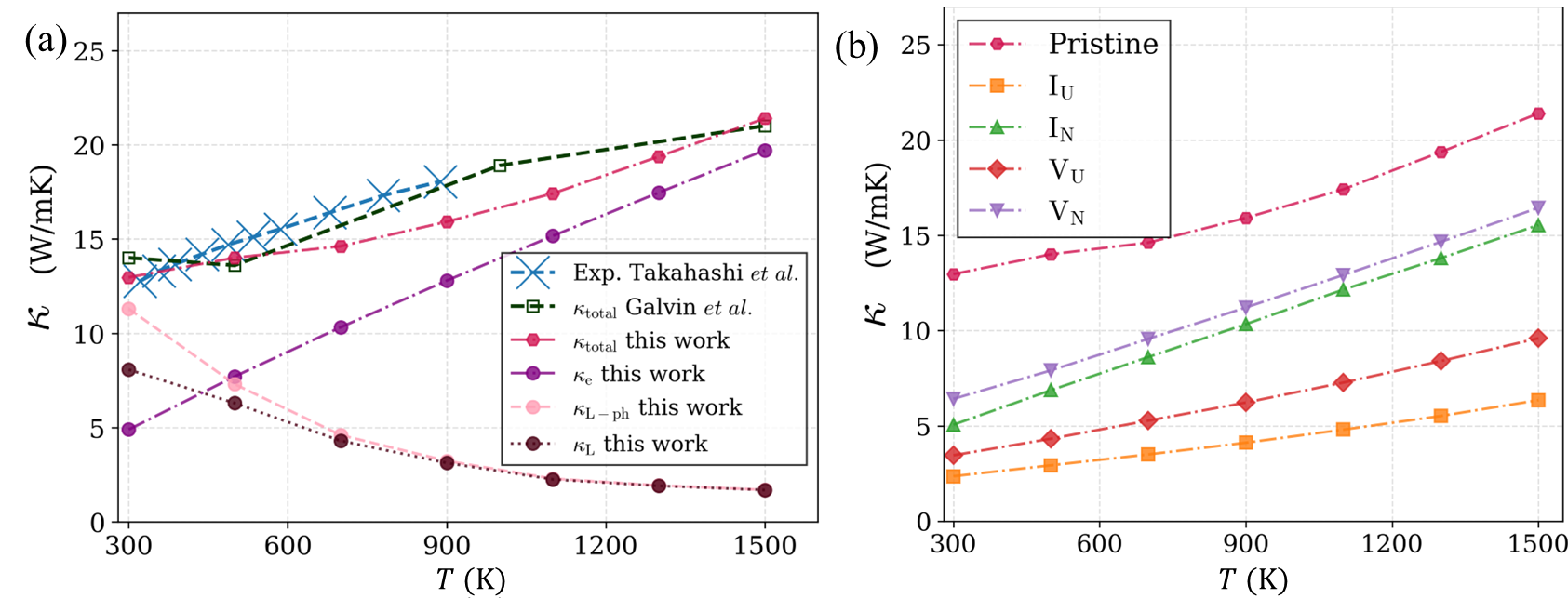}
    \caption{(a) $\kappa_\mathrm{total}$, $\kappa_\mathrm{L-ph}$, $\kappa_\mathrm{e}$, and $\kappa_\mathrm{L}$ calculated in this work for pristine UN, compared with experimental data from Takahashi \textit{et al.} \cite{Takahashi1971} and previously reported $\kappa_\mathrm{total}$ values from Galvin \textit{et al.} \cite{Galvin2025}, where electron--phonon coupling was not included. (b) $\kappa_\mathrm{total}$ for pristine UN and UN containing 0.46\% point defects: $\mathrm{I}_\mathrm{U}$, $\mathrm{I}_\mathrm{N}$, $\mathrm{V}_\mathrm{U}$, and $\mathrm{V}_\mathrm{N}$. The electronic thermal conductivity $\kappa_\mathrm{e}$ is estimated from the Wiedemann–Franz law combined with the semi-classical electrical resistivity model following Zhou \textit{et al.} \cite{Zhou2020,Zhou2021}.}
    \label{fig:5}
\end{figure}

To obtain the total thermal conductivity of UN, the phonon relaxation time ($\tau_\mathrm{ph}$) derived from the NMA method was corrected to account for electron--phonon (e--ph) scattering, yielding $\tau$ and ultimately $\kappa_\mathrm{L}$. The estimated electronic thermal conductivity, $\kappa_\mathrm{e}$ was then added to obtain the total thermal conductivity, $\kappa_\mathrm{total}$ ($\kappa_\mathrm{L}$ +  $\kappa_\mathrm{e}$).  The relative roles of $\kappa_\mathrm{L}$ and $\kappa_\mathrm{e}$  vary continuously with temperature. From cryogenic temperatures up to room temperature, $\kappa_\mathrm{L}$ is dominant \cite{Chen2025} and is mainly determined by three-phonon scattering and electron–phonon coupling effect. In the range $300~\mathrm{K} < T < 600~\mathrm{K}$, both $\kappa_\mathrm{L}$ and $\kappa_\mathrm{e}$ make significant contributions. For $T > 600~\mathrm{K}$, $\kappa_\mathrm{e}$ becomes increasingly important with rising temperature, and four-phonon scattering effect for $\kappa_\mathrm{L}$ becomes important in this temperature range. Fig.~\ref{fig:5}(a) shows $\kappa_\mathrm{L-ph}$, $\kappa_\mathrm{L}$,  $\kappa_\mathrm{e}$ and $\kappa_\mathrm{total}$ for pristine UN over the temperature range 300--1500 K, together with experimental data from Takahashi \textit{et al.} \cite{Takahashi1971} and previous computational results \cite{Galvin2025}. The inclusion of electron--phonon coupling significantly reduces $\kappa_\mathrm{L}$ at low temperatures. As temperature increases, this effect diminishes rapidly and becomes negligible above $\sim$600 K, where intrinsic phonon--phonon scattering dominates the lattice thermal resistance.
At 300 K, our $\kappa_\mathrm{total}$ predictions demonstrate excellent agreement with experimental measurements. This highlights the critical necessity of incorporating both the electron-phonon (e-ph) coupling resistance and the purely electronic contribution ($\kappa_\mathrm{e}$). Prior MD studies, such as those by Galvin \textit{et al.} \cite{Galvin2025}, show an overestimation of $\kappa_\mathrm{total}$ at room temperature due to the neglect of the e-ph coupling. With increasing temperature, the intrinsic phonon-phonon interactions become the dominant source of thermal resistance with minimal e-ph coupling effect. In the high-temperature regime, both the $\kappa_\mathrm{total}$ values calculated in this work and those from Galvin \textit{et al.} \cite{Galvin2025} exhibit comparable values for the pristine system.

\begin{figure}
    \centering
    \includegraphics[width=0.9\linewidth]{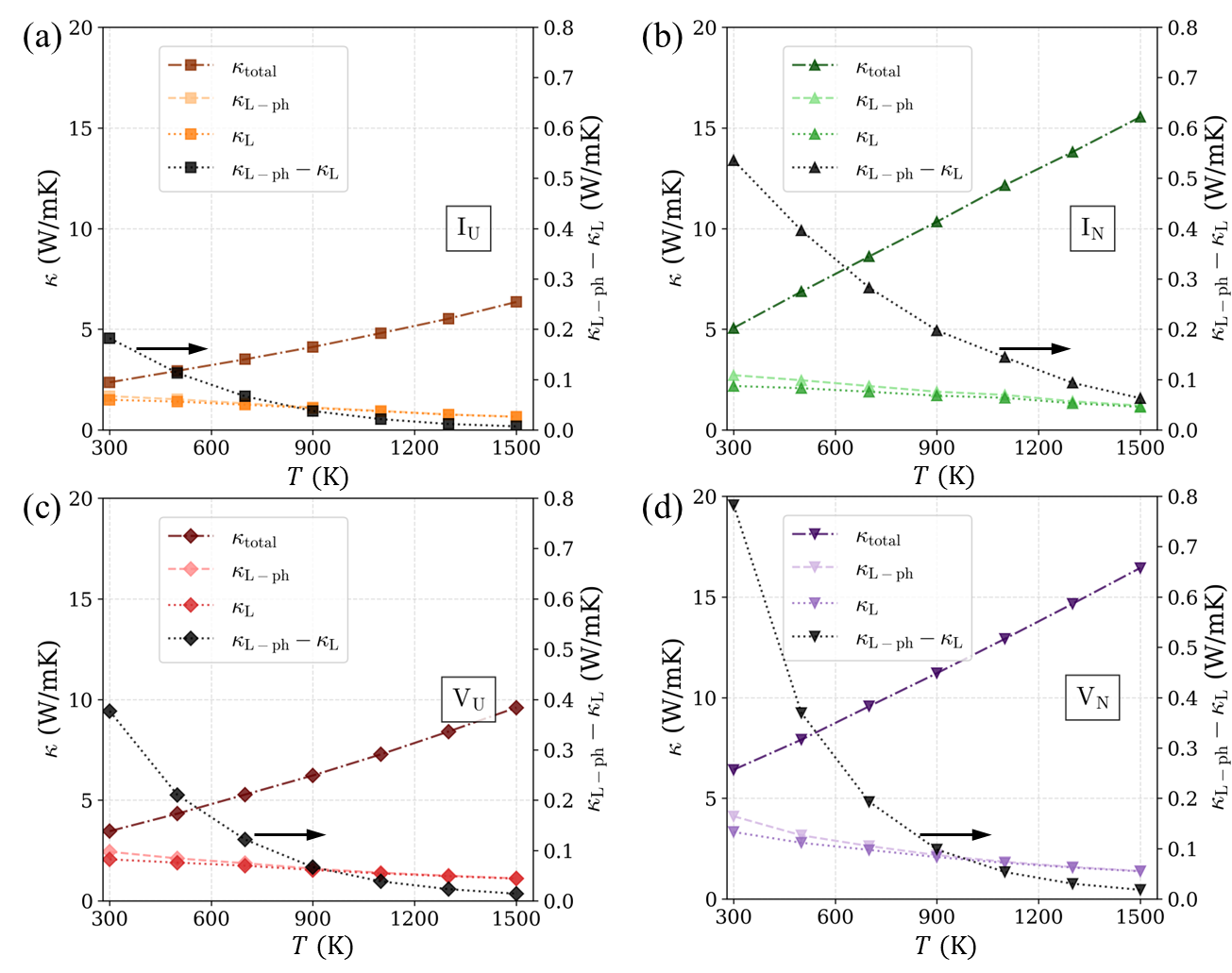}
    \caption{$\kappa_\mathrm{total}$, $\kappa_\mathrm{L-ph}$, $\kappa_\mathrm{L}$ (left y-axis) and $\kappa_\mathrm{L-ph}$-$\kappa_\mathrm{L}$ (right y-axis) for UN containing 0.46\% point defects as a function of temperature: (a) $\mathrm{I}_\mathrm{U}$, (b) $\mathrm{I}_\mathrm{N}$, (c) $\mathrm{V}_\mathrm{U}$, and (d) $\mathrm{V}_\mathrm{N}$, respectively.}
    \label{fig:6}
\end{figure}

The values for $\kappa_\mathrm{L-ph}$, $\kappa_\mathrm{L}$ and $\kappa_\mathrm{total}$ for UN containing 0.46\% $\mathrm{I}_\mathrm{U}$, $\mathrm{I}_\mathrm{N}$, $\mathrm{V}_\mathrm{U}$, and $\mathrm{V}_\mathrm{N}$ are shown in Fig.~\ref{fig:6}(a-d), respectively. Notably, while electron-phonon coupling still exerts a noticeable effect ($\kappa_\mathrm{L-ph}$-$\kappa_\mathrm{L}$ in Fig.~\ref{fig:6} (b) and (d)) on UN with nitrogen-related defects at $300$ K, its impact is almost entirely negligible for UN with uranium-related defects across all investigated temperatures. The $\kappa_\mathrm{total}$ for pristine and defective UN is summarized in Fig.~\ref{fig:5}(b). The degradation impact of point defects on total thermal conductivity follows the sequence $\mathrm{I}_\mathrm{U} > \mathrm{V}_\mathrm{U} > \mathrm{I}_\mathrm{N} > \mathrm{V}_\mathrm{N}$, which is consistent with the hierarchy observed in the $\kappa_\mathrm{L-ph}$ results from the NMA method. The major loss in thermal conductivity due to defect-phonon scattering is compensated by the electronic contribution. At elevated temperatures, as $\kappa_\mathrm{L-ph}$ converges to similar values (Fig.~\ref{fig:kL_defect}(b)), the wide difference in $\kappa_\mathrm{total}$ between the various defect types is mainly due to the defect-specific degradation of $\kappa_\mathrm{e}$ with the trend: $\mathrm{I}_\mathrm{U} >\mathrm{V}_\mathrm{U} >\mathrm{I}_\mathrm{N} > \mathrm{V}_\mathrm{N}$ within the present semi-classical estimate. 

It is worth emphasizing that the estimation of $\kappa_\mathrm{e}$ in defective UN deserves future methodological improvements. For phonon transport, MLIP combined with lattice dynamics and direct molecular-dynamics formalisms provides an effective route in which anharmonic and defect-induced scattering can be properly treated. For electronic transport in actinide systems with point defects, it remains challenging to precisely quantify the contribution. First-principles studies based on DFT and the BTE~\cite{Kocevski2022,Kocevski2023,Szpunar2014} have addressed pristine UN, where the principal challenge is the treatment of the strongly correlated U $5f$ electrons. As Yin \textit{et al.}~\cite{Yin2011} have shown for related actinide systems, $5f$ correlation modifies not only electron--electron interactions but also electron--phonon coupling, so that transport coefficients depend on the level of theory used to describe the correlated electronic structure (DFT$+U$, dynamical mean-field theory, or hybrid functionals). Extending these treatments to defective supercells, where each defect introduces localized states that can hybridize with the $5f$ states near the Fermi level, substantially increases the computational complexity. The complementary approach, used for example by Galvin \textit{et al.}~\cite{Galvin2025} via BoltzTraP~\cite{Madsen2018}, evaluates the defect-induced change in the electronic density of states; this captures one contribution to $\rho$ but does not include electron--defect scattering, which is the dominant defect contribution at the concentrations considered here.

The semiclassical resistivity framework adopted in this work, following Zhou \textit{et al.}~\cite{Zhou2020,Zhou2021}, restores the electron--defect scattering channel through $\rho_\mathrm{e\text{-}pd}$. This treatment represents a compromise, and its limitations arise from how it represents electron--defect scattering. The scattering cross section is approximated as a geometric quantity, $A_\mathrm{cs}=\pi r^{2}$, derived from the atomic radius~\cite{Slater1964}; this treats the defect as a hard-sphere scatterer whose strength is set by atomic size rather than by the matrix elements between Bloch states and the perturbing defect potential~\cite{mahan2013many}. As a result, the framework cannot distinguish between defects of the same species when they induce different local electrostatic potentials and different defect-related electronic states near $E_\mathrm{F}$: the difference between $\mathrm{I}_\mathrm{U}$ and $\mathrm{V}_\mathrm{U}$, or between $\mathrm{I}_\mathrm{N}$ and $\mathrm{V}_\mathrm{N}$, enters only through the defect volume $V_\mathrm{pd}$, and the predicted ordering $\mathrm{I}_\mathrm{U} > \mathrm{V}_\mathrm{U} > \mathrm{I}_\mathrm{N} > \mathrm{V}_\mathrm{N}$ reflects primarily the contrast between U and N atomic radii. The Fermi energy is taken from a pristine-UN DFT calculation and is held fixed across all defective cases, consistent with its role in Eq.~\eqref{rho_general} as setting the carrier velocity of the host conduction band, which at the dilute defect concentration considered here ($x=0.46\%$) should be well approximated by the pristine value. In reality, the concentration of these defects depends on irradiation conditions, which may be inferred from recent experimental results using rate theory modeling and often proves to be smaller \cite{KAMBOJ2026117322,HE2021116778}. Furthermore, the electron relaxation time $\tau_\mathrm{e}$ is calibrated against pristine-UN resistivity and applied uniformly to the defective cases, though defects are expected to shorten $\tau_\mathrm{e}$ through additional scattering channels. Bounding the resulting error would require a self-consistent treatment of the defect-modified electronic structure and the corresponding scattering rate, which is not currently available. Future development in this direction may benefit from combining correlated-electron methods such as DFT+DMFT or DFT$+U$ for the electronic structure of defective supercells, electron--phonon Wannier-interpolation frameworks to evaluate e--ph coupling in a consistent manner.

\section{Conclusion}
This work clarifies how point defect type, anharmonicity, and electronic transport collectively govern thermal conductivity in UN. The analysis shows that defect-induced degradation is strongly influenced by defect nature and by how each defect interacts with different regions of the phonon spectrum. Uranium interstitials introduce strong, broadband scattering that suppresses heat-carrying modes, whereas other point defects produce more selective and mode-dependent effects, including reduced scattering for certain low-frequency phonons. These differences lead to nontrivial changes in the relative contributions of acoustic and optical phonons. The comparison between Green--Kubo and normal mode analysis further highlights that, in defective systems, thermal transport cannot be fully described within a simple phonon-based framework. In particular, the results demonstrate that electronic transport plays a dominant role at elevated temperatures ($>\sim 600$ K) and electron-phonon coupling becomes negligible in reducing conductivity in defective systems. These findings provide a basis for understanding thermal transport in UN under irradiation conditions. Looking forward, a more accurate treatment of $\kappa_\mathrm{e}$ in defective UN, in particular a first-principles description of electron–defect scattering that accounts for $5f$ correlation effects on both electron–electron and electron–phonon interactions is necessary.

\section*{Supplementary Material}
Supporting information as noted in the manuscript.

\section*{Acknowledgments}
This work is supported by the Center for Thermal Energy Transport under Irradiation, an Energy Frontier Research Center funded by the U.S. Department of Energy, Office of Science, Office of Basic Energy Sciences.

\section*{DATA AVAILABILITY}
Data available within the article and its supplementary materials

\bibliographystyle{unsrt}
\bibliography{reference}

@book{mahan2013many,
  title={Many-particle physics},
  author={Mahan, Gerald D},
  year={2013},
  publisher={Springer Science \& Business Media}
}

@article{Samsel,
   author = {M. Samsel-Czekała and E. Talik and P. de V. Du Plessis and R. Troć and H. Misiorek and C. Sułkowski},
   doi = {10.1103/PhysRevB.76.144426},
   issn = {1098-0121},
   issue = {14},
   journal = {Physical Review B},
   month = {10},
   pages = {144426},
   title = {Electronic structure and magnetic and transport properties of single-crystalline {UN}},
   volume = {76},
   url = {https://link.aps.org/doi/10.1103/PhysRevB.76.144426},
   year = {2007}
}

@article{hanus2021thermal,
  title={Thermal transport in defective and disordered materials},
  author={Hanus, Riley and Gurunathan, Ramya and Lindsay, Lucas and Agne, Matthias T and Shi, Jingjing and Graham, Samuel and Jeffrey Snyder, G},
  journal={Applied physics reviews},
  volume={8},
  number={3},
  year={2021},
  publisher={AIP Publishing}
}

@article{tamura1983isotope,
  title={Isotope scattering of dispersive phonons in {Ge}},
  author={Tamura, Shin-ichiro},
  journal={Physical Review B},
  volume={27},
  number={2},
  pages={858},
  year={1983},
  publisher={APS}
}

@article{thomas2010predicting,
  title={Predicting phonon dispersion relations and lifetimes from the spectral energy density},
  author={Thomas, John A and Turney, Joseph E and Iutzi, Ryan M and Amon, Cristina H and McGaughey, Alan JH},
  journal={Physical Review B—Condensed Matter and Materials Physics},
  volume={81},
  number={8},
  pages={081411},
  year={2010},
  publisher={APS}
}

@article{lv2016direct,
  title={Direct calculation of modal contributions to thermal conductivity via Green--Kubo modal analysis},
  author={Lv, Wei and Henry, Asegun},
  journal={New Journal of Physics},
  volume={18},
  number={1},
  pages={013028},
  year={2016},
  publisher={IOP Publishing}
}

@article{allen1989thermal,
  title={Thermal conductivity of glasses: Theory and application to amorphous {Si}},
  author={Allen, Philip B and Feldman, Joseph L},
  journal={Physical review letters},
  volume={62},
  number={6},
  pages={645},
  year={1989},
  publisher={APS}
}

@article{simoncelli2019unified,
  title={Unified theory of thermal transport in crystals and glasses},
  author={Simoncelli, Michele and Marzari, Nicola and Mauri, Francesco},
  journal={Nature Physics},
  volume={15},
  number={8},
  pages={809--813},
  year={2019},
  publisher={Nature Publishing Group UK London}
}

@article{Ross1988,
   author = {Steven B. Ross and Mohamed S. El-Genk and R. Bruce Matthews},
   doi = {10.1016/0022-3115(88)90026-8},
   issn = {0022-3115},
   issue = {3},
   journal = {Journal of Nuclear Materials},
   month = {2},
   pages = {318-326},
   publisher = {North-Holland},
   title = {Thermal conductivity correlation for uranium nitride fuel between 10 and 1923 {K}},
   volume = {151},
   year = {1988},
}

@article{Szpunar2020,
   author = {Barbara Szpunar and Jayangani I. Ranasinghe and Linu Malakkal and Jerzy A. Szpunar},
   doi = {10.1016/j.jpcs.2020.109636},
   issn = {00223697},
   journal = {Journal of Physics and Chemistry of Solids},
   month = {11},
   pages = {109636},
   title = {First principles investigation of thermal transport of uranium mononitride},
   volume = {146},
   year = {2020},
}

@article{AbdulHameed2024,
   author = {Mohamed AbdulHameed and Benjamin Beeler and Conor O.T. Galvin and Michael W.D. Cooper},
   doi = {10.1016/J.JNUCMAT.2024.155247},
   issn = {0022-3115},
   journal = {Journal of Nuclear Materials},
   month = {6},
   pages = {155247},
   publisher = {North-Holland},
   title = {Assessment of uranium nitride interatomic potentials},
   year = {2024},
}

@article{Kocevski2022,
   author = {Vancho Kocevski and Michael W.D. Cooper and Antoine J. Claisse and David A. Andersson},
   doi = {10.1016/j.jnucmat.2022.153553},
   issn = {00223115},
   journal = {Journal of Nuclear Materials},
   month = {4},
   pages = {153553},
   title = {Development and application of a uranium mononitride ({UN}) potential: Thermomechanical properties and {Xe} diffusion},
   volume = {562},
   year = {2022},
}

@article{Kocevski2023,
   author = {Vancho Kocevski and Daniel A. Rehn and Adrien J. Terricabras and Arjen van Veelen and Michael W.D. Cooper and Scarlett Widgeon Paisner and Sven C. Vogel and Joshua T. White and David A. Andersson},
   doi = {10.1016/j.jnucmat.2023.154241},
   issn = {00223115},
   journal = {Journal of Nuclear Materials},
   month = {4},
   pages = {154241},
   title = {Finite temperature properties of uranium mononitride},
   volume = {576},
   url = {https://linkinghub.elsevier.com/retrieve/pii/S0022311523000119},
   year = {2023}
}

@article{Takahashi1971,
   author = {Y. Takahashi and M. Murabayashi and Y. Akimoto and T. Mukaibo},
   doi = {10.1016/0022-3115(71)90059-6},
   issn = {00223115},
   issue = {3},
   journal = {Journal of Nuclear Materials},
   month = {3},
   pages = {303-308},
   title = {Uranium mononitride: Heat capacity and thermal conductivity from 298 to 1000 °K},
   volume = {38},
   url = {https://linkinghub.elsevier.com/retrieve/pii/0022311571900596},
   year = {1971},
}

@article{Chen2025,
   author = {Beihan Chen and Zilong Hua and Jennifer K. Watkins and Linu Malakkal and Marat Khafizov and David H. Hurley and Miaomiao Jin},
   doi = {10.1063/5.0294389},
   issn = {0021-8979},
   issue = {20},
   journal = {Journal of Applied Physics},
   month = {11},
   title = {Machine learning interatomic potential for predicting the thermal properties of uranium nitride},
   volume = {138},
   url = {https://pubs.aip.org/jap/article/138/20/205102/3373278/Machine-learning-interatomic-potential-for},
   year = {2025}
}

@article{Galvin2025,
   author = {C.O.T. Galvin and A. Schneider and M. AbdulHameed and B. Beeler and R.W. Grimes},
   doi = {10.1016/j.pnucene.2025.105923},
   issn = {01491970},
   journal = {Progress in Nuclear Energy},
   month = {11},
   pages = {105923},
   title = {Atomic-scale modeling assessing the impact of defects on the thermal conductivity of {UN}},
   volume = {189},
   year = {2025}
}

@article{Lan2025,
   author = {Yulin Lan and Tianhao Rui and Zhuangzhuang Ma and Linyuan Lu and Yunhao Wang and Yang Yu and Mingxuan Deng and Tianxing Lan and Zhekang Zhao and Junjie Wang and Congyi Li and Haibin Zhang},
   doi = {10.3390/CRYST15050459},
   issn = {2073-4352},
   issue = {5},
   journal = {Crystals 2025, Vol. 15, Page 459},
   month = {5},
   pages = {459},
   publisher = {Multidisciplinary Digital Publishing Institute},
   title = {First-Principles Investigation of the Effect of Vacancy Defects and Carbon Impurities on Thermal Conductivity of Uranium Mononitride ({UN})},
   volume = {15},
   url = {https://www.mdpi.com/2073-4352/15/5/459/htm https://www.mdpi.com/2073-4352/15/5/459},
   year = {2025}
}

@article{Green1954,
   author = {Melville S. Green},
   doi = {10.1063/1.1740082},
   issn = {00219606},
   issue = {3},
   journal = {The Journal of Chemical Physics},
   title = {Markoff random processes and the statistical mechanics of time-dependent phenomena. II. Irreversible processes in fluids},
   volume = {22},
   year = {1954},
}

@article{Kubo1957,
   author = {Ryogo Kubo},
   doi = {10.1143/JPSJ.12.570},
   issn = {13474073},
   issue = {6},
   journal = {Journal of the Physical Society of Japan},
   title = {Statistical'Mechanical Theory of Irreversible Processes. I. General Theory and Simple Applications to Magnetic and Conduction Problems},
   volume = {12},
   year = {1957},
}

@article{McGaughey2004,
   author = {A. J. H. McGaughey and M. Kaviany},
   doi = {10.1103/PhysRevB.69.094303},
   issn = {1098-0121},
   issue = {9},
   journal = {Physical Review B},
   month = {3},
   pages = {094303},
   title = {Quantitative validation of the {Boltzmann} transport equation phonon thermal conductivity model under the single-mode relaxation time approximation},
   volume = {69},
   year = {2004},
}

@article{Turney2009,
   author = {J. E. Turney and E. S. Landry and A. J. H. McGaughey and C. H. Amon},
   doi = {10.1103/PhysRevB.79.064301},
   issn = {1098-0121},
   issue = {6},
   journal = {Physical Review B},
   month = {2},
   pages = {064301},
   title = {Predicting phonon properties and thermal conductivity from anharmonic lattice dynamics calculations and molecular dynamics simulations},
   volume = {79},
   year = {2009},
}

@techReport{Speidel1963,
   author = {E.O. Speidel and D.L. Keller},
   doi = {10.2172/4674236},
   institution = {Battelle Memorial Institute (United States)},
   month = {5},
   title = {FABRICATION AND PROPERTIES OF HOT-PRESSED URANIUM MONONITRIDE},
   url = {http://www.osti.gov/servlets/purl/4674236-X8ym0X/},
   year = {1963}
}

@techreport{Endebrock1964,
  author       = {Endebrock, R. W. and Foster Jr., E. L. and Keller, D. L.},
  title        = {Preparation and Properties of Cast {UN}},
  institution  = {Battelle Memorial Institute},
  address      = {Columbus, Ohio},
  number       = {BMI-1690; EURAEC-1206},
  year         = {1964}
}

@article{Hayes1990tc,
   author = {S.L. Hayes and J.K. Thomas and K.L. Peddicord},
   doi = {10.1016/0022-3115(90)90376-X},
   issn = {00223115},
   issue = {2-3},
   journal = {Journal of Nuclear Materials},
   month = {5},
   pages = {289-299},
   title = {Material property correlations for uranium mononitride: {III}. Transport properties},
   volume = {171},
   url = {https://linkinghub.elsevier.com/retrieve/pii/002231159090376X},
   year = {1990},
}

@article{Jones2023,
   author = {Suzanne Jones and Colin Boxall and Chris Maher and Robin Taylor},
   doi = {10.1016/J.PNUCENE.2023.104917},
   issn = {0149-1970},
   journal = {Progress in Nuclear Energy},
   month = {11},
   pages = {104917},
   publisher = {Pergamon},
   title = {A review of the reprocessability of uranium nitride based fuels},
   volume = {165},
   year = {2023}
}

@article{Youinou2014,
   author = {Gilles J. Youinou and R. Sonat Sen},
   doi = {10.13182/NT14-22},
   issn = {19437471},
   issue = {2},
   journal = {Nuclear Technology},
   month = {11},
   pages = {123-138},
   publisher = {Taylor & Francis},
   title = {Impact of Accident-Tolerant Fuels and Claddings on the Overall Fuel Cycle: A Preliminary Systems Analysis},
   volume = {188},
   url = {https://www.tandfonline.com/doi/abs/10.13182/NT14-22},
   year = {2014}
}

@article{MOORE1970,
   author = {J. P. MOORE and W. FULKERSON and D. L. McELROY},
   doi = {10.1111/j.1151-2916.1970.tb12014.x},
   issn = {00027820},
   issue = {2},
   journal = {Journal of the American Ceramic Society},
   month = {2},
   pages = {76-82},
   title = {Thermal Conductivity, Electrical Resistivity, and Seebeck Coefficient of Uranium Mononitride},
   volume = {53},
   url = {https://onlinelibrary.wiley.com/doi/10.1111/j.1151-2916.1970.tb12014.x},
   year = {1970}
}

@article{NASU1968,
   author = {Shōichi NASU and Takeo KIKUCHI},
   doi = {10.1080/18811248.1968.9732463},
   issn = {0022-3131},
   issue = {6},
   journal = {Journal of Nuclear Science and Technology},
   month = {6},
   pages = {318-319},
   title = {Thermal Diffusivity of {UN} from 20° to 1,000{°C} by Laser Pulse Method},
   volume = {5},
   url = {http://www.tandfonline.com/doi/abs/10.1080/18811248.1968.9732463},
   year = {1968}
}

@article{Kikuchi1973,
   author = {T. Kikuchi and T. Takahashi and S. Nasu},
   doi = {10.1016/0022-3115(73)90162-1},
   issn = {00223115},
   issue = {4},
   journal = {Journal of Nuclear Materials},
   month = {1},
   pages = {284-292},
   title = {Porosity dependence of thermal conductivity of uranium mononitride},
   volume = {45},
   year = {1973}
}

@article{Ladd1986,
   author = {Anthony J. C. Ladd and Bill Moran and William G. Hoover},
   doi = {10.1103/PhysRevB.34.5058},
   issn = {0163-1829},
   issue = {8},
   journal = {Physical Review B},
   month = {10},
   pages = {5058-5064},
   title = {Lattice thermal conductivity: A comparison of molecular dynamics and anharmonic lattice dynamics},
   volume = {34},
   year = {1986},
}

@book{srivastava2022physics,
  title={The physics of phonons},
  author={Srivastava, Gyaneshwar P},
  year={2022},
  publisher={CRC press}
}

@article{Kurosaki2000,
   author = {Ken Kurosaki and Kimihiko Yano and Kazuhiro Yamada and Masayoshi Uno and Shinsuke Yamanaka},
   doi = {10.1016/S0925-8388(00)01127-0},
   issn = {09258388},
   issue = {2},
   journal = {Journal of Alloys and Compounds},
   month = {10},
   pages = {305-310},
   title = {A molecular dynamics study of the thermal conductivity of uranium mononitride},
   volume = {311},
   url = {https://linkinghub.elsevier.com/retrieve/pii/S0925838800011270},
   year = {2000}
}

@article{PORTER199753,
title = {Atomistic modeling of finite-temperature properties of {$\beta$-SiC}. {I. Lattice} vibrations, heat capacity, and thermal expansion},
journal = {Journal of Nuclear Materials},
volume = {246},
number = {1},
pages = {53-59},
year = {1997},
issn = {0022-3115},
doi = {10.1016/S0022-3115(97)00035-4},
author = {Lisa J. Porter and Ju Li and Sidney Yip}
}

@incollection{Klemens1958thermal,
  title={Thermal conductivity and lattice vibrational modes},
  author={Klemens, PG},
  booktitle={Solid state physics},
  volume={7},
  pages={1--98},
  year={1958},
  publisher={Elsevier}
}

@article{Klemens1955,
doi = {10.1088/0370-1298/68/12/303},
year = {1955},
month = {dec},
publisher = {},
volume = {68},
number = {12},
pages = {1113},
author = {P G Klemens},
title = {The Scattering of Low-Frequency Lattice Waves by Static Imperfections},
journal = {Proceedings of the Physical Society. Section A}
}

@article{Callaway1959,
   author = {Joseph Callaway},
   doi = {10.1103/PhysRev.113.1046},
   issn = {0031-899X},
   issue = {4},
   journal = {Physical Review},
   month = {2},
   pages = {1046-1051},
   title = {Model for Lattice Thermal Conductivity at Low Temperatures},
   volume = {113},
   year = {1959},
}

@article{Walker1963,
   author = {C. T. Walker and R. O. Pohl},
   doi = {10.1103/PhysRev.131.1433},
   issn = {0031-899X},
   issue = {4},
   journal = {Physical Review},
   month = {8},
   pages = {1433-1442},
   title = {Phonon Scattering by Point Defects},
   volume = {131},
   url = {https://link.aps.org/doi/10.1103/PhysRev.131.1433},
   year = {1963}
}

@article{Feng2017,
   author = {Tianli Feng and Lucas Lindsay and Xiulin Ruan},
   doi = {10.1103/PhysRevB.96.161201},
   issn = {2469-9950},
   issue = {16},
   journal = {Physical Review B},
   month = {10},
   pages = {161201},
   title = {Four-phonon scattering significantly reduces intrinsic thermal conductivity of solids},
   volume = {96},
   url = {https://link.aps.org/doi/10.1103/PhysRevB.96.161201},
   year = {2017}
}

@Article{LAMMPS,
  author = "A. P. Thompson and H. M. Aktulga and R. Berger and 
     D. S. Bolintineanu and W. M. Brown and P. S. Crozier and
     P. J. in 't Veld and A. Kohlmeyer and S. G. Moore and T. D. Nguyen and
     R. Shan and M. J. Stevens and J. Tranchida and C. Trott and S. J. Plimpton",
  title = "{LAMMPS} - a flexible simulation tool for
     particle-based materials modeling at the 
     atomic, meso, and continuum scales",
  journal = "Comp. Phys. Comm.",
  volume =  "271",
  pages =   "108171",
  year =    "2022",
  doi = "10.1016/j.cpc.2021.108171"
}

@article{Henry2008,
   author = {Asegun S. Henry and Gang Chen},
   doi = {10.1166/jctn.2008.2454},
   issn = {15461955},
   issue = {2},
   journal = {Journal of Computational and Theoretical Nanoscience},
   title = {Spectral phonon transport properties of silicon based on molecular dynamics simulations and lattice dynamics},
   volume = {5},
   year = {2008},
}

@article{Madsen2018,
   author = {Georg K.H. Madsen and Jesús Carrete and Matthieu J. Verstraete},
   doi = {10.1016/j.cpc.2018.05.010},
   issn = {00104655},
   journal = {Computer Physics Communications},
   month = {10},
   pages = {140-145},
   title = {{BoltzTraP2}, a program for interpolating band structures and calculating semi-classical transport coefficients},
   volume = {231},
   url = {https://linkinghub.elsevier.com/retrieve/pii/S0010465518301632},
   year = {2018}
}

@article{Resnick2019,
   author = {Alex Resnick and Katherine Mitchell and Jungkyu Park and Eduardo B. Farfán and Tien Yee},
   doi = {10.1016/j.net.2019.03.011},
   issn = {17385733},
   issue = {5},
   journal = {Nuclear Engineering and Technology},
   month = {8},
   pages = {1398-1405},
   title = {Thermal transport study in actinide oxides with point defects},
   volume = {51},
   url = {https://linkinghub.elsevier.com/retrieve/pii/S1738573318307812},
   year = {2019}
}

@article{Hurley2022,
   author = {David H. Hurley and Anter El-Azab and Matthew S. Bryan and Michael W. D. Cooper and Cody A. Dennett and Krzysztof Gofryk and Lingfeng He and Marat Khafizov and Gerard H. Lander and Michael E. Manley and J. Matthew Mann and Chris A. Marianetti and Karl Rickert and Farida A. Selim and Michael R. Tonks and Janelle P. Wharry},
   doi = {10.1021/acs.chemrev.1c00262},
   issn = {0009-2665},
   issue = {3},
   journal = {Chemical Reviews},
   month = {2},
   pages = {3711-3762},
   publisher = {American Chemical Society},
   title = {Thermal Energy Transport in Oxide Nuclear Fuel},
   volume = {122},
   year = {2022},
}

@article{Mitchell2020,
   author = {Katherine Mitchell and Jungkyu Park and Alex Resnick and Hunter Horner and Eduardo B. Farfan},
   doi = {10.3390/app10051860},
   issn = {2076-3417},
   issue = {5},
   journal = {Applied Sciences},
   month = {3},
   pages = {1860},
   title = {Phonon Scattering and Thermal Conductivity of Actinide Oxides with Defects},
   volume = {10},
   year = {2020}
}

@article{Watanabe2009,
   author = {Taku Watanabe and Srinivasan G. Srivilliputhur and Patrick K. Schelling and James S. Tulenko and Susan B. Sinnott and Simon R. Phillpot},
   doi = {10.1111/j.1551-2916.2009.02966.x},
   issn = {0002-7820},
   issue = {4},
   journal = {Journal of the American Ceramic Society},
   month = {4},
   pages = {850-856},
   title = {Thermal Transport in Off‐Stoichiometric Uranium Dioxide by Atomic Level Simulation},
   volume = {92},
   year = {2009}
}

@article{Liu2016,
   author = {X.-Y. Liu and M.W.D. Cooper and K.J. McClellan and J.C. Lashley and D.D. Byler and B.D.C. Bell and R.W. Grimes and C.R. Stanek and D.A. Andersson},
   doi = {10.1103/PhysRevApplied.6.044015},
   issn = {2331-7019},
   issue = {4},
   journal = {Physical Review Applied},
   month = {10},
   pages = {044015},
   title = {Molecular Dynamics Simulation of Thermal Transport in {UO$_2$}  Containing Uranium, Oxygen, and Fission-product Defects},
   volume = {6},
   url = {https://link.aps.org/doi/10.1103/PhysRevApplied.6.044015},
   year = {2016}
}

@article{Yang2025,
   author = {Jiahui Yang and Yandong Sun and Ben Xu},
   doi = {10.1103/PhysRevB.111.104112},
   issn = {2469-9950},
   issue = {10},
   journal = {Physical Review B},
   month = {3},
   pages = {104112},
   title = {Impact of point defects on the thermal conductivity of {GaN} studied using machine-learned potentials},
   volume = {111},
   year = {2025}
}

@article{Li2025,
   author = {Guotai Li and Zheng Cui and Ruiqiang Guo},
   doi = {10.1103/PhysRevB.111.094102},
   issn = {2469-9950},
   issue = {9},
   journal = {Physical Review B},
   month = {3},
   pages = {094102},
   title = {Impact of point defect charge states on the thermal conductivity of {AlN}},
   volume = {111},
   year = {2025}
}

@article{Rounds2018,
   author = {Robert Rounds and Biplab Sarkar and Dorian Alden and Qiang Guo and Andrew Klump and Carsten Hartmann and Toru Nagashima and Ronny Kirste and Alexander Franke and Matthias Bickermann and Yoshinao Kumagai and Zlatko Sitar and Ramón Collazo},
   doi = {10.1063/1.5028141},
   issn = {0021-8979},
   issue = {18},
   journal = {Journal of Applied Physics},
   month = {5},
   title = {The influence of point defects on the thermal conductivity of {AlN} crystals},
   volume = {123},
   url = {https://pubs.aip.org/jap/article/123/18/185107/154826/The-influence-of-point-defects-on-the-thermal},
   year = {2018}
}

@article{Han2022,
   author = {Zherui Han and Xiaolong Yang and Wu Li and Tianli Feng and Xiulin Ruan},
   doi = {10.1016/j.cpc.2021.108179},
   issn = {00104655},
   journal = {Computer Physics Communications},
   month = {1},
   pages = {108179},
   title = {FourPhonon: An extension module to {ShengBTE} for computing four-phonon scattering rates and thermal conductivity},
   volume = {270},
   url = {https://linkinghub.elsevier.com/retrieve/pii/S0010465521002915},
   year = {2022}
}

@article{McGaughey2013,
   author = {Alan J. H. McGaughey and Jason M. Larkin},
   doi = {10.1615/annualrevheattransfer.2013006915},
   issn = {1049-0787},
   issue = {N/A},
   journal = {Annual Review of Heat Transfer},
   title = {PREDICTING PHONON PROPERTIES FROM EQUILIBRIUM MOLECULAR DYNAMICS SIMULATIONS},
   volume = {17},
   year = {2013}
}

@article{phono3py,
  title = {Distributions of phonon lifetimes in Brillouin zones},
  author = {Togo, Atsushi and Chaput, Laurent and Tanaka, Isao},
  journal = {Phys. Rev. B},
  volume = {91},
  issue = {9},
  pages = {094306},
  numpages = {31},
  year = {2015},
  month = {Mar},
  publisher = {American Physical Society},
  doi = {10.1103/PhysRevB.91.094306},
}

@article{phonopy-3py,
  author  = {Togo, Atsushi and Chaput, Laurent and Tadano, Terumasa and Tanaka, Isao},
  title   = {Implementation strategies in phonopy and phono3py},
  journal = {J. Phys. Condens. Matter},
  volume  = {35},
  number  = {35},
  pages   = {353001},
  year    = {2023},
  doi     = {10.1088/1361-648X/acd831}
}

@article{phonopy,
  author  = {Togo, Atsushi},
  title   = {First-principles Phonon Calculations with Phonopy and Phono3py},
  journal = {J. Phys. Soc. Jpn.},
  volume  = {92},
  number  = {1},
  pages   = {012001},
  year    = {2023},
  doi     = {10.7566/JPSJ.92.012001}
}

@article{Kuksin2016,
   author = {A. Yu Kuksin and S. V. Starikov and D. E. Smirnova and V. I. Tseplyaev},
   doi = {10.1016/J.JALLCOM.2015.10.223},
   issn = {0925-8388},
   journal = {Journal of Alloys and Compounds},
   month = {2},
   pages = {385-394},
   publisher = {Elsevier},
   title = {The diffusion of point defects in uranium mononitride: {Combination of DFT and atomistic simulation with novel potential}},
   volume = {658},
   year = {2016}
}

@article{Mishin2005,
   author = {Y. Mishin and M.J. Mehl and D.A. Papaconstantopoulos},
   doi = {10.1016/j.actamat.2005.05.001},
   issn = {13596454},
   issue = {15},
   journal = {Acta Materialia},
   month = {9},
   pages = {4029-4041},
   title = {Phase stability in the {Fe–Ni} system: Investigation by first-principles calculations and atomistic simulations},
   volume = {53},
   year = {2005}
}

@article{Szpunar2014,
   author = {B. Szpunar and J. A. Szpunar},
   doi = {10.1155/2014/178360},
   issn = {2356-7066},
   journal = {International Journal of Nuclear Energy},
   month = {9},
   pages = {1-7},
   title = {Thermal Conductivity of Uranium Nitride and Carbide},
   volume = {2014},
   url = {https://www.hindawi.com/journals/ijne/2014/178360/},
   year = {2014}
}

@book{Ziman2001,
   author = {John M Ziman},
   publisher = {Oxford university press},
   title = {Electrons and phonons: the theory of transport phenomena in solids},
   year = {2001}
}

@article{Tong2019,
   author = {Zhen Tong and Shouhang Li and Xiulin Ruan and Hua Bao},
   doi = {10.1103/PhysRevB.100.144306},
   issn = {2469-9950},
   issue = {14},
   journal = {Physical Review B},
   month = {10},
   pages = {144306},
   title = {Comprehensive first-principles analysis of phonon thermal conductivity and electron-phonon coupling in different metals},
   volume = {100},
   url = {https://link.aps.org/doi/10.1103/PhysRevB.100.144306},
   year = {2019}
}

@article{Pang2013,
   author = {Judy W. L. Pang and William J. L. Buyers and Aleksandr Chernatynskiy and Mark D. Lumsden and Bennett C. Larson and Simon R. Phillpot},
   doi = {10.1103/PhysRevLett.110.157401},
   issn = {0031-9007},
   issue = {15},
   journal = {Physical Review Letters},
   month = {4},
   pages = {157401},
   title = {Phonon Lifetime Investigation of Anharmonicity and Thermal Conductivity of   {UO$_2$}   by Neutron Scattering and Theory},
   volume = {110},
   url = {https://link.aps.org/doi/10.1103/PhysRevLett.110.157401},
   year = {2013}
}

@article{Zhou2021,
   author = {Shuxiang Zhou and Yongfeng Zhang and Dane Morgan},
   doi = {10.1016/j.jnucmat.2021.153044},
   issn = {00223115},
   journal = {Journal of Nuclear Materials},
   month = {9},
   publisher = {Elsevier B.V.},
   title = {An ab-initio based semi-empirical thermal conductivity model for multiphase uranium-zirconium alloys},
   volume = {553},
   year = {2021}
}

@misc{Zhou2020,
   author = {Shuxiang Zhou},
   month = {4},
   title = {First Principle based Thermal Conductivity Modeling of Metal Alloys with Applications on Uranium Alloys},
   url = {https://search.library.wisc.edu/digital/AHTXDVBFKNPQRO9C},
   year = {2020}
}

@article{Slater1964,
   author = {J. C. Slater},
   doi = {10.1063/1.1725697},
   isbn = {202601:06:25},
   issn = {0021-9606},
   issue = {10},
   journal = {The Journal of Chemical Physics},
   month = {11},
   pages = {3199-3204},
   publisher = {AIP Publishing},
   title = {Atomic Radii in Crystals},
   volume = {41},
   url = {/aip/jcp/article/41/10/3199/352896/Atomic-Radii-in-Crystals},
   year = {1964}
}

@article{Yin2011,
   author = {Quan Yin and Andrey Kutepov and Kristjan Haule and Gabriel Kotliar and Sergey Y. Savrasov and Warren E. Pickett},
   doi = {10.1103/PhysRevB.84.195111},
   issn = {1098-0121},
   issue = {19},
   journal = {Physical Review B},
   month = {11},
   pages = {195111},
   title = {Electronic correlation and transport properties of nuclear fuel materials},
   volume = {84},
   url = {https://link.aps.org/doi/10.1103/PhysRevB.84.195111},
   year = {2011}
}

@misc{Miller2025,
   author = {Zachary Miller and Massimiliano Fratoni and Alex Levinsky and Galen  T. Craven and Christopher Matthews and Joshua  T. White and Anders  David Ragnar Andersson and Maria Kosmidou and Adrien  J. Terricabras},
   doi = {10.2139/SSRN.5417748},
   title = {Thermo-mechanical property correlations for uranium mononitride with uncertainty quantification},
   url = {https://papers.ssrn.com/abstract=5417748},
   year = {2025}
}

@misc{Valter2015,
   author = {Mikael Valter},
   title = {Thermal Conductivity of Uranium Mononitride},
   url = {https://urn.kb.se/resolve?urn=urn:nbn:se:liu:diva-122337},
   year = {2015}
}

@misc{Miller2024,
   author = {Zachary Aaron Miller and Alex Levinsky and Galen Thomas Craven and Vedant Kiritkumar Mehta and Massimiliano Fratoni and Joshua Taylor White and Anders David Ragnar Andersson and Maria Kosmidou and Adrien Jose Emile Terricabras},
   city = {Los Alamos, NM (United States)},
   doi = {10.2172/2440180},
   institution = {Los Alamos National Laboratory (LANL)},
   month = {9},
   title = {Uranium Mononitride (UN) Handbook},
   url = {https://www.osti.gov/servlets/purl/2440180/},
   year = {2024}
}

@article{Charatsidou2026Zr,
    title = {Impact of zirconium incorporation on the thermophysical properties of uranium mononitride},
    journal = {Journal of Nuclear Materials},
    volume = {623},
    pages = {156467},
    year = {2026},
    issn = {0022-3115},
    doi = {https://doi.org/10.1016/j.jnucmat.2026.156467},
    url = {https://www.sciencedirect.com/science/article/pii/S0022311526000334},
    author = {Elina Charatsidou and Anita Pazzaglia and Kaitlyn Bullock and Maria Giamouridou and Eleanor Lawrence Bright and Mikael Jolkkonen and Christoph Hennig and Pär Olsson}
}

@misc{Charatsidou2026,
   author = {Elina Charatsidou and Angus  P. C. Wylie and Nils Wikström and Maria Giamouridou and Riley Moeykens and Robert  J.W. Frost and Ericmoore Jossou and Michael Short and Pär Olsson},
   month = {2},
   title = {Thermal Diffusivity of Ion-Irradiated Uranium Mononitride},
   url = {https://papers.ssrn.com/abstract=6459442},
   year = {2026}
}

@article{KAMBOJ2026117322,
title = {Excessive dislocation loop growth in Uranium Mononitride under high temperature proton irradiation},
journal = {Scripta Materialia},
volume = {280},
pages = {117322},
year = {2026},
issn = {1359-6462},
doi = {https://doi.org/10.1016/j.scriptamat.2026.117322},
url = {https://www.sciencedirect.com/science/article/pii/S1359646226001582},
author = {Anshul Kamboj and Lin-Chieh Yu and Md Minaruzzaman and Kaustubh Bawane and Zilong Hua and Lin Shao and Marat Khafizov and Yongfeng Zhang and Miaomiao Jin and Jennifer K. Watkins and Chao Jiang and David H. Hurley and Boopathy Kombaiah}
}

@article{HE2021116778,
title = {Phase and defect evolution in uranium-nitrogen-oxygen system under irradiation},
journal = {Acta Materialia},
volume = {208},
pages = {116778},
year = {2021},
issn = {1359-6454},
doi = {https://doi.org/10.1016/j.actamat.2021.116778},
url = {https://www.sciencedirect.com/science/article/pii/S1359645421001580},
author = {Lingfeng He and Marat Khafizov and Chao Jiang and Beata Tyburska-Püschel and Brian J. Jaques and Pengyuan Xiu and Peng Xu and Mitchell K. Meyer and Kumar Sridharan and Darryl P. Butt and Jian Gan}
}

@article{Kresse1996,
   author = {G. Kresse and J. Furthmüller},
   doi = {10.1016/0927-0256(96)00008-0},
   issn = {09270256},
   issue = {1},
   journal = {Computational Materials Science},
   month = {7},
   pages = {15-50},
   title = {Efficiency of ab-initio total energy calculations for metals and semiconductors using a plane-wave basis set},
   volume = {6},
   year = {1996}
}

@article{Kresse1996i,
   author = {G. Kresse and J. Furthmüller},
   doi = {10.1103/PhysRevB.54.11169},
   issn = {0163-1829},
   issue = {16},
   journal = {Physical Review B},
   month = {10},
   pages = {11169-11186},
   title = {Efficient iterative schemes for ab initio total-energy calculations using a plane-wave basis set},
   volume = {54},
   year = {1996}
}

@article{RYU2020109615,
title = {Comparison and validation of the lattice thermal conductivity formulas used in equilibrium molecular dynamics simulations for binary systems},
journal = {Computational Materials Science},
volume = {178},
pages = {109615},
year = {2020},
issn = {0927-0256},
doi = {https://doi.org/10.1016/j.commatsci.2020.109615},
url = {https://www.sciencedirect.com/science/article/pii/S0927025620301063},
author = {Jinho Ryu and Takuji Oda and Hisashi Tanigawa},
}

@article{Manjunatha2021,
  author  = {Manjunatha, L. and Takamatsu, H. and Cannon, J. J.},
  title   = {Atomic-level Breakdown of Green--Kubo Relations Provides New Insight into the Mechanisms of Thermal Conduction},
  journal = {Scientific Reports},
  volume  = {11},
  number  = {1},
  pages   = {5597},
  year    = {2021},
  doi     = {10.1038/s41598-021-84446-9},
  url     = {https://doi.org/10.1038/s41598-021-84446-9},
  publisher = {Nature Publishing Group}
}

@article{Gurunathan2020,
   author = {Ramya Gurunathan and Riley Hanus and Maxwell Dylla and Ankita Katre and G. Jeffrey Snyder},
   doi = {10.1103/PhysRevApplied.13.034011},
   issn = {2331-7019},
   issue = {3},
   journal = {Physical Review Applied},
   month = {3},
   pages = {034011},
   title = {Analytical Models of Phonon–Point-Defect Scattering},
   volume = {13},
   url = {https://link.aps.org/doi/10.1103/PhysRevApplied.13.034011},
   year = {2020}
}

\end{document}


\title{Supplementary Materials for: `` Thermal Transport in Defective Uranium Nitride: Effects of Point Defects, Anharmonicity, and Electronic Contributions " }

\author{Beihan Chen}
 \affiliation{Department of Nuclear Engineering, The Pennsylvania State University, University Park, PA 16802, USA}
 
\author{Marat Khafizov}
\affiliation{Department of Mechanical and Aerospace Engineering, The Ohio State University, Columbus, OH 43210, USA}

\author{Zilong Hua}
\affiliation{Idaho National Laboratory, Idaho Falls, ID 83415, USA}

\author{David H. Hurley}
\affiliation{Idaho National Laboratory, Idaho Falls, ID 83415, USA}

\author{Miaomiao Jin}%
 \email{mmjin@psu.edu}
\affiliation{Department of Nuclear Engineering, The Pennsylvania State University, University Park, PA 16802, USA}%

\maketitle

\renewcommand{\thefigure}{S\arabic{figure}}
\renewcommand{\thesection}{\arabic{section}}

\section{The construction of defective supercells}


For the defect concentration of 0.46\%, the conventional unit cell containing 8 atoms was expanded to a $3\times3\times3$ supercell with 216 atoms. One U or N atom was then removed from lattice position or added at a tetrahedral site, resulting in a defect concentration of 0.46\% (1/216). This $3\times3\times3$ supercell was subsequently expanded by $2\times2\times2$ to form a $6\times6\times6$ supercell with point defects at the same concentration, corresponding to 0.46\% (8/1728).

\section{An example of the integration of lattice thermal conductivity using the GK method}

\begin{figure}[htbp]
    \centering
    \includegraphics[width=1.0\linewidth]{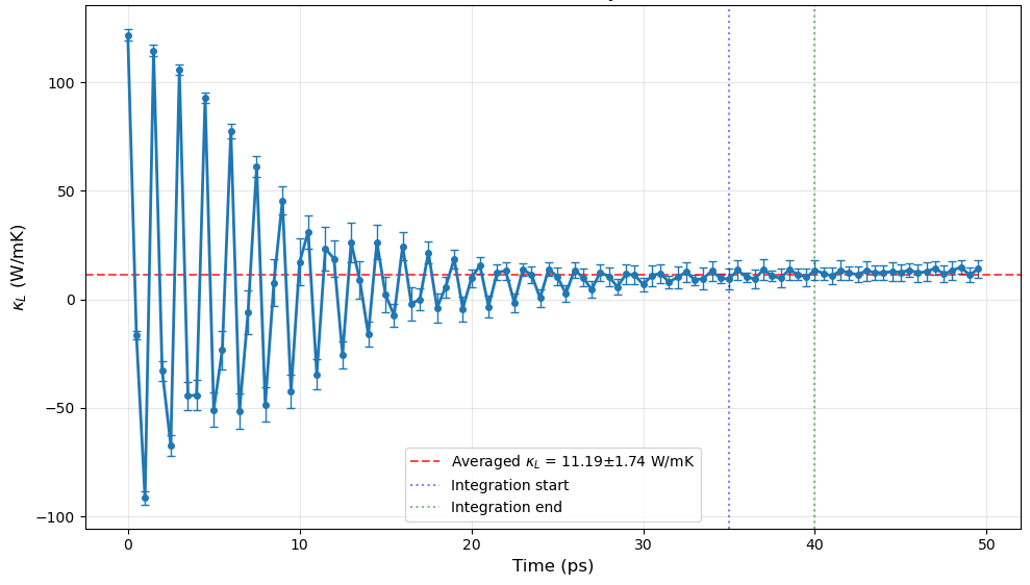}
    \caption{An example showing the convergence of lattice thermal conductivity calculated using the Green--Kubo method. The thermal conductivity approaches a stable value as the integration time increases.}
    \vspace{0.5em}
    {\textbf{Alt text:} An example of the convergence of $\kappa_\mathrm{L}$ integrated using the GK method.}
    \label{fig:kappa}
\end{figure}

Figure \ref{fig:kappa} illustrates the convergence of the lattice thermal conductivity ($\kappa_\mathrm{L}$) as a function of integration time. The curve shown corresponds to the ensemble average of ten independent simulations, each initiated with a distinct random seed. For improved clarity, the raw data are averaged over 0.5 ps intervals. To reduce the influence of long-tail noise and statistical fluctuations, the final reported $\kappa_\mathrm{L}$ is obtained using a 5 ps stationary averaging window applied after the integration reaches a converged plateau (shown between the integration start dash line and integration end dash line).

\section{An example of the fitting of phonon relaxation time using the NMA method}

\begin{figure}[htbp]
    \centering
    \includegraphics[width=0.75\linewidth]{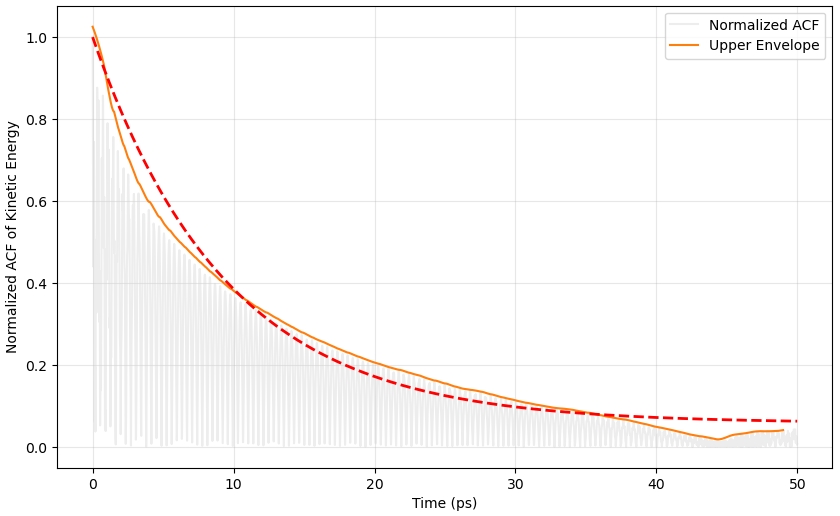}
    \caption{Normalized modal energy autocorrelation function of a representative phonon mode in pristine UN, averaged over ten independent molecular dynamics simulations. The solid curve "Upper Envelope" shows the fit using the damped oscillatory function of Eq. (7), from which the phonon relaxation time ($\tau_n$) is extracted. The oscillation amplitude decays exponentially while the fitted autocorrelation function remains oscillatory. }
    \label{fig:tau}
    \vspace{0.5em}
    {\textbf{Alt text:} An example of the fit of $\tau_n$ for pristine UN. }
\end{figure}

Figure \ref{fig:tau} shows a representative normalized modal energy autocorrelation function for pristine UN, averaged over ten independent molecular dynamics simulations. The autocorrelation function is fitted using the damped oscillatory expression given in Eq. (7),
\begin{equation}
    \mathrm{exp}(-\frac{t}{\tau_{n-\mathrm{ph}}})\mathrm{cos}^{2}(\omega_{n} t) 
\end{equation}

where the exponential term describes the decay of the oscillation amplitude and the fitting parameter ($\tau_n$) is taken as the phonon relaxation time.

\begin{figure}[htbp]
    \centering
    \includegraphics[width=0.75\linewidth]{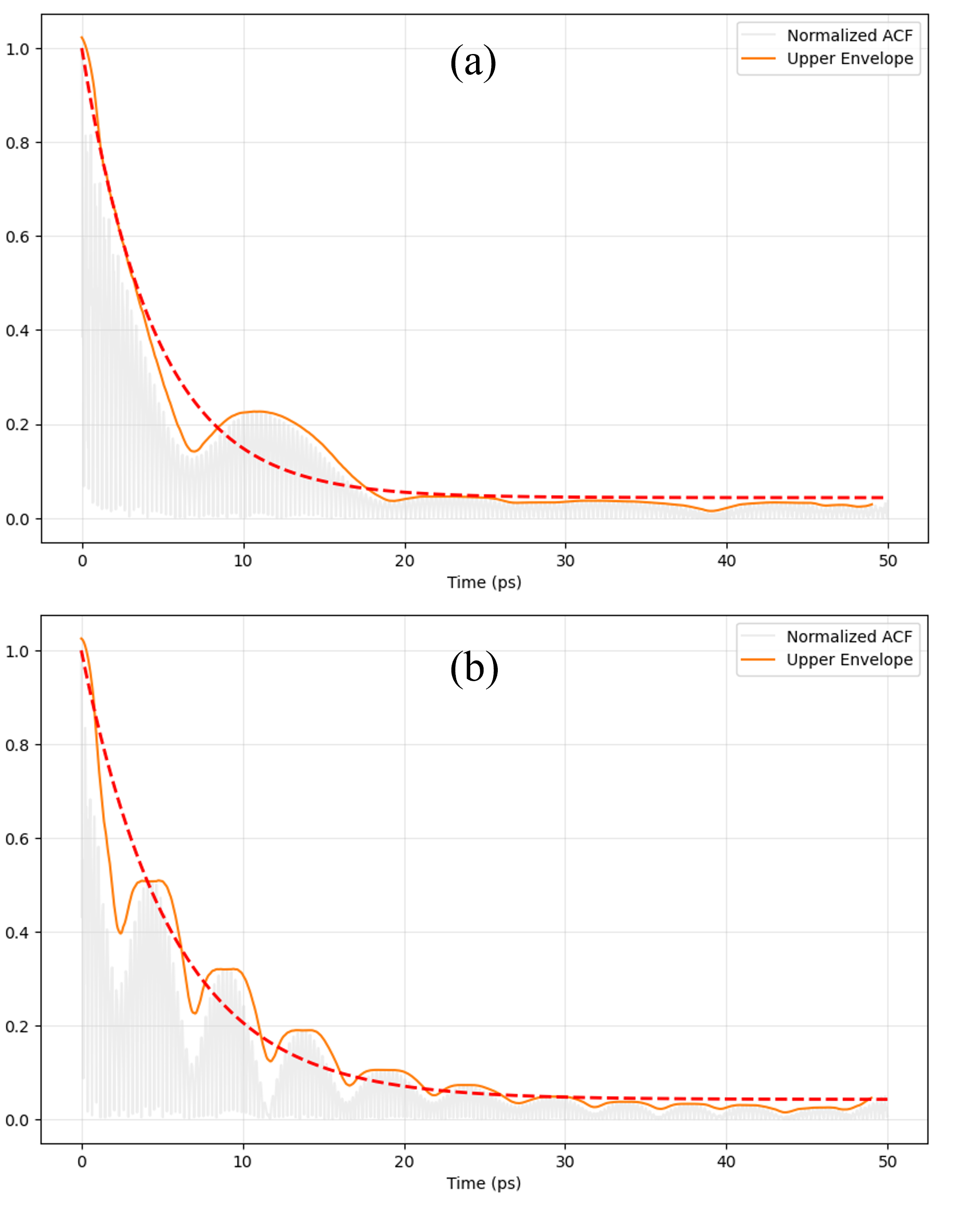}
    \caption{Line plots of representative normalized modal energy autocorrelation functions for defective uranium nitride. Compared with the pristine case, the oscillations exhibit stronger fluctuations and less regular decay. Damped oscillatory fitting curves capture the overall trend but deviate from portions of the data, illustrating the increased uncertainty in extracting effective phonon relaxation times in the presence of point defects.}
    {\textbf{Alt text:} Two examples of the fit of $\tau_n$ from UN containing $\mathrm{I}_\mathrm{N}$.}
    \label{fig:tau_defect}
\end{figure}

Representative results for defective UN are shown in Fig. \ref{fig:tau_defect}. Compared with pristine UN, the autocorrelation functions exhibit substantially stronger fluctuations and more complex decay behavior due to defect-induced disorder and enhanced mode scattering. Consequently, the damped oscillatory model provides only an effective description of the decay, and the extracted relaxation times should be interpreted as effective lifetimes. The fitting becomes more sensitive to statistical noise and the selected fitting window, contributing to the larger uncertainty in the NMA-derived lattice thermal conductivity for defective systems.

\section{ Impact of $\mathrm{I}_\mathrm{N}$ }

\begin{figure}
    \centering
    \includegraphics[width=0.6\linewidth]{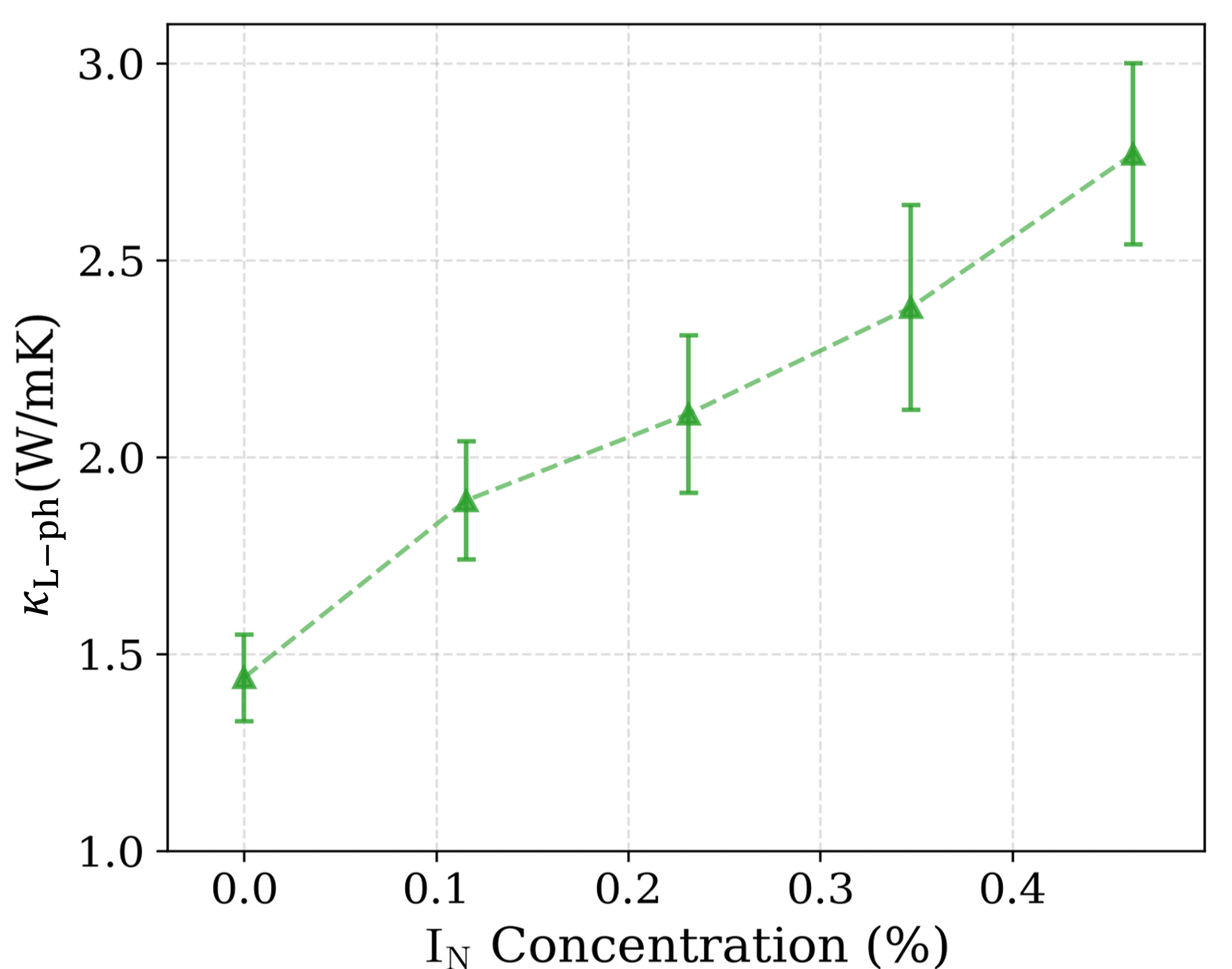}
    \caption{ GK calculated $\kappa_\mathrm{L-ph}$ at 1500 K for UN system containing $\mathrm{I}_\mathrm{N}$ at different concentrations, where 0.0 \% corresponds to bulk UN. }
    {\textbf{Alt text:} $\kappa_\mathrm{L-ph}$ at 1500 K versus $\mathrm{I}_\mathrm{N}$ concentrations.}
    \label{fig:I_N}
\end{figure}

\begin{figure}
    \centering
    \includegraphics[width=0.9\linewidth]{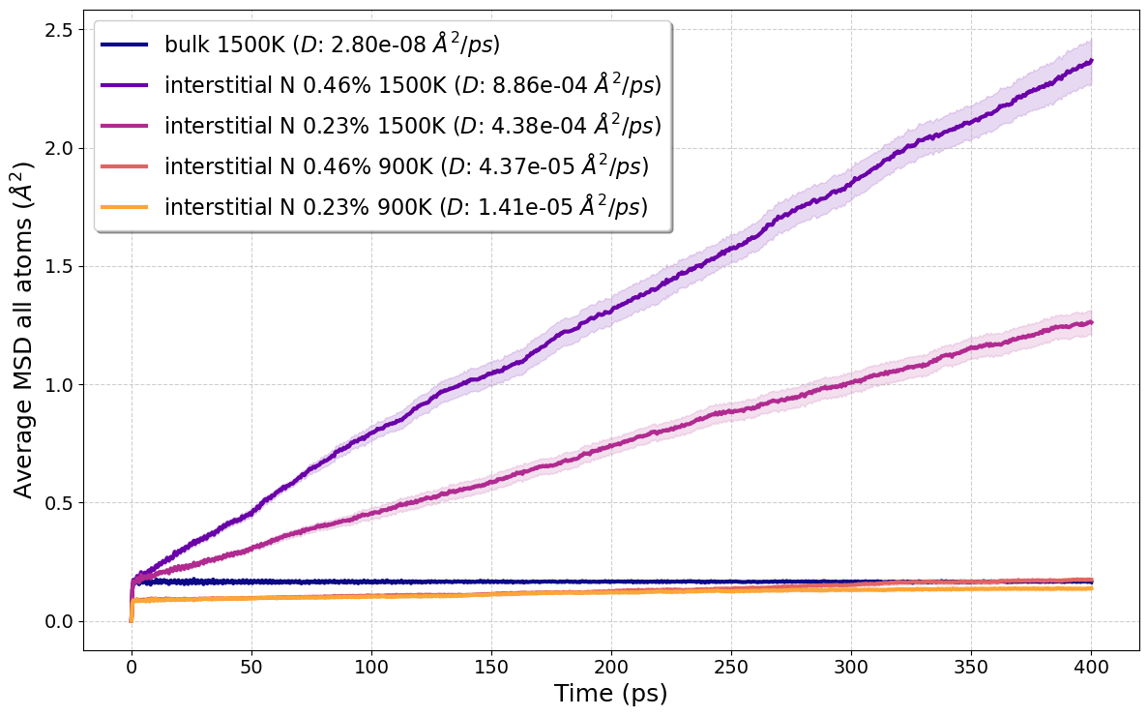}
    \caption{ MSD at 900 K and 1500 K for UN system containing $\mathrm{I}_\mathrm{N}$ at different concentrations. Compared to pristine UN, $\mathrm{I}_\mathrm{N}$ shows no mobility at 900 K, while at 1500 K it gives different trends. }
    {\textbf{Alt text:} MSD of UN with $\mathrm{I}_\mathrm{N}$ at different concentrations compared with pristine UN at 900 K and 1500 K.}
    \label{fig:msd}
\end{figure}


The impact of $\mathrm{I}_\mathrm{N}$ on the GK $\kappa_\mathrm{L-ph}$ at 1500 K is evaluated and shown in Fig.~\ref{fig:I_N}. The increase in GK $\kappa_\mathrm{L-ph}$ with $\mathrm{I}_\mathrm{N}$ concentration can be understood from Fig.~\ref{fig:msd}, which shows that the average MSD of the UN system containing $\mathrm{I}_\mathrm{N}$ increases noticeably with time at 1500 K, at a rate proportional to the $\mathrm{I}_\mathrm{N}$ concentration, indicating translation of the center of mass. In contrast, atoms in pristine UN exhibit almost no motion at 1500 K, and at 900 K the MSD of UN containing $\mathrm{I}_\mathrm{N}$ remains small, close to that of pristine UN.